\pdfoutput=1 %Allow PDF figures
\documentclass[twocolumn, tighten]{aastex6}
\usepackage{longtable}
\usepackage{hyperref}
\usepackage{afterpage}

\newcommand{\degrees}{$^\circ$}
\newcommand{\arcminutes}{\arcmin{}}
\newcommand{\arcseconds}{\farcs{}}
\newcommand{\hours}{\textsuperscript{h}}
\newcommand{\minutes}{\textsuperscript{m}}
\newcommand{\seconds}{\fs{}}
\newcommand{\htwo}{H$_2$} %Molecular hydrogen abbreviation
\newcommand{\hnaught}{H$^0$} %Atomic hydrogen abbreviation
\newcommand{\hplus}{H$^+$} %Ionized hydrogen abbreviation
\newcommand{\rotation}{$J$} %Symbol for rotation or rotation level
\newcommand{\vibration}{$v$} %Symbol for vibration or vibrational level
\newcommand{\sectionsymbol}{$\S$} %Symbol for section

\begin{document}

\title{Excitation of Molecular Hydrogen in the Orion Bar Photodissociation Region
from
a Deep Near-Infrared IGRINS Spectrum}
\author{Kyle F. Kaplan\altaffilmark{1}\email{kfkaplan@astro.as.utexas.edu}}
\author{Harriet L. Dinerstein\altaffilmark{1}\email{harriet@astro.as.utexas.edu}}
\author{Heeyoung Oh\altaffilmark{2}\altaffilmark{3}\email{hyoh@astro.snu.ac.kr}}
\author{Gregory N. Mace\altaffilmark{1}\email{gmace@astro.as.utexas.edu}}
\author{Hwihyun Kim\altaffilmark{4}\altaffilmark{2}\altaffilmark{1}\email{hkim@gemini.edu}}
\author{Kimberly R. Sokal\altaffilmark{1}\email{kimberly.sokal@astro.as.utexas.edu}}
\author{Michael D. Pavel\altaffilmark{1}\email{michaeldpavel@gmail.com}}
\author{Sungho Lee\altaffilmark{2}\email{leesh@kasi.re.kr}}
\author{Soojong Pak\altaffilmark{5}\email{soojong@khu.ac.kr}}
\author{Chan Park\altaffilmark{2}\email{chanpark@kasi.re.kr}}
\author{Jae Sok Oh\altaffilmark{2}\email{ojs001@kasi.re.kr}}
\author{Daniel T. Jaffe\altaffilmark{1}\email{dtj@astro.as.utexas.edu}}
\altaffiltext{1}{Department of Astronomy, University of Texas at Austin, 2515 Speedway, Stop C1400, Austin, TX 78712-1205, USA.}
\altaffiltext{2}{Korea Astronomy and Space Science Institute, 776 Daedeok-daero, Yuseong-gu, Daejeon 34055, Korea.}
\altaffiltext{3}{Korea University of Science and Technology, 217 Gajeong-ro, Yuseong-gu, Daejeon 34113, Korea.}
\altaffiltext{4}{Gemini Observatory, Southern Operations Center, c/o AURA, Casilla 603, La Serena, Chile.}
\altaffiltext{5}{School of Space Research and Institute of Natural Sciences, Kyung Hee University, 1732 Deogyeong-daero, Giheung-gu, Yongin-si, Gyeonggi-do 17104, Korea.}

\begin{abstract}
We present a deep near-infrared spectrum of the Orion Bar Photodissociation Region (PDR) taken with the Immersion Grating INfrared Spectrometer (IGRINS) on the 2.7 m telescope at the McDonald Observatory.  IGRINS has high spectral resolution ($R\sim45000$) and instantaneous broad wavelength coverage (1.45--2.45 $\mu$m), enabling us to detect
87
emission lines from rovibrationally excited molecular hydrogen (\htwo{}) that arise from transitions out of 69 upper rovibration levels of the electronic ground state.  These levels cover a large range of rotational and vibrational quantum numbers and excitation energies, making them 
excellent probes
of the excitation mechanisms of \htwo{} and physical conditions within the PDR.  
The Orion Bar PDR is thought to consist of cooler high density clumps or filaments 
($T=50$ to $250$~K, $n_H = 10^5$ to $10^7$~cm$^{-3}$)
embedded in a warmer lower density medium
($T=250$ to $1000$~K,  $n_H=10^4$ to  $10^5$~cm$^{-3}$).
We fit a grid of
constant temperature and density
Cloudy models, which recreate the observed \htwo{} level populations well, to constrain the temperature to a range of 
600 to 650~K and the density to $n_H = 2.5\times10^3$ to $10^4$~cm$^{-3}$.
The 
best-fit
model gives $T = 625$~K and $n_H = 5 \times 10^3$~cm$^{-3}$.
This
well-constrained
warm temperature is consistent with kinetic temperatures found by other studies for the Orion Bar's lower density medium.
However, the range of densities well fit by the model grid is
marginally
lower than those reported by other studies.  We could be  observing lower density gas than the surrounding medium, or perhaps a density-sensitive parameter in our models is not properly estimated.\end{abstract}


\section{Introduction}
Ultraviolet (UV) radiation regulates the process of star formation and the energetics, ionization state, and chemistry of the interstellar medium (ISM).  Photodissociation or Photon-dominated Regions (PDRs) are regions in the ISM at the 
interface between hot ionized gas and cool molecular gas that are energetically dominated by 
non-ionizing UV photons.  PDRs arise around regions of massive star formation or star death \citep{tielens85, hollenbach97, hollenbach99} and make up the bulk of the neutral ISM in star forming galaxies such as the Milky Way.
In the most extreme cases, starburst galaxies can have much of their starlight reprocessed and reradiated by PDRs.  

The canonical model for a PDR, as presented by \citet{tielens85}, is a plane-parallel slab of gas  illuminated on one side by stellar UV radiation.  The interaction between UV photons and the gas sets up a differentiated structure
that
can be characterized by the phases of hydrogen, which transition from predominantly ionized  (\hplus{}), to neutral atomic (\hnaught{}), and then to molecular (\htwo{}).
The \hplus{}/\hnaught{} and  \hnaught{}/\htwo{} interfaces are called the ``ionization'' and ``dissociation'' fronts respectively.
Extreme-UV (EUV) photons with energies above the \hnaught{} ionization potential ($> 13.6$~eV, $\lambda < 912$~\AA{}) pass through the \hplus{} zone and are absorbed by \hnaught{} at the ionization front.  Far-UV (FUV) photons with energies just below the Lyman continuum 
($11.2$--$13.6$~eV,
$912 <  \lambda < 1110$~\AA{}) pass through the \hnaught{} zone but are attenuated by dust, elements with lower ionization potentials, and the Lyman and Werner bands of \htwo{}.  Beyond the dissociation front, the remaining FUV photons are rapidly absorbed as the cloud transitions to cold molecular gas.

The \htwo{} rotational (\rotation{}) and vibrational (\vibration{}), hereafter referred to as ``rovibrational,''  levels of the ground electronic state \citep{black76} can be excited by  two main processes:  UV excitation (fluorescence) and collisional (thermal) excitation. UV excitation occurs when FUV photons absorbed by \htwo{} excite the molecules to upper electronic states (through the Lyman and Werner bands), from which $\sim 10\%$ of the molecules will dissociate \citep{field66}, and the rest decay into bound excited rovibrational levels \citep{black76, black87}.  Collisional excitation occurs
via inelastic collisions with other particles in the gas that leave
the molecules in the excited rovibrational levels of the ground electronic state.  These rovibrationally excited molecules decay via either collisions or a radiative cascade of rovibrational transitions, which have wavelengths ranging from the mid-infrared to the optical.
\htwo{} is a homonuclear diatomic molecule lacking a permanent electronic dipole moment,
so the rovibrational transitions
occur only as electric quadrupole transitions ($\Delta~J=~0,~\pm2$), which are optically thin under most conditions.
Since the line flux from each transition is proportional to the number of molecules in an upper level, observing many lines allows us to calculated the detailed \htwo{} rovibrational level populations.
These emission lines have significant diagnostic power to reveal physical conditions within PDRs at the \hnaught{}/\htwo{} dissociation front where most of the emission arises.

The  UV and collisional excitation and de-excitation processes give rise to two limiting cases for the rovibrational level populations: one that is thermal, and the other for pure UV radiative excitation.  In dense and/or hot gas (such as in shocks), the rovibrational levels are excited and de-excited by frequent collisions and the level populations approach a thermal distribution.  In 
low-density
cool gas exposed to UV radiation, the level populations take on a distinctive non-thermal distribution that does not decline monotonically with increasing excitation energy.  Observations of the rovibrational line flux ratios readily distinguish between these two limiting cases.
However, in practice, many sources show observed level populations intermediate between UV excited and thermal \htwo{}.  Two possible reasons are the superposition of spatially unresolved components, or collisional modification of UV excited \htwo{}.
\citet{sternberg89} and \citet{burton90} show that dense PDRs can exhibit level populations that are modified from the pure UV excited case by collisions.  
Collisions easily dominate the excitation of \htwo{} into states with low energies above the ground, bringing these states into thermal equilibrium with the gas.   States at higher energies are primarily UV excited but collisional de-excitation modifies their populations from the pure UV excited case.  This collisional modification of the level populations in UV excited \htwo{} can mimic the spectrum from an unresolved combination of pure UV excited and thermal components.

Dense interstellar PDRs are found in star forming regions where molecular clouds are exposed to UV radiation from newly formed hot massive stars. 
At a distance of $\sim400-500$~pc \citep{schlafly14}, the Orion Nebula is the nearest example of such a high-mass region, and it serves as an archetype for the more distant star forming regions found elsewhere in the Milky Way and in starburst galaxies.
The optically visible part of the Orion Nebula is an  \hplus{}
(or \ion{H}{2})
region where the
massive OB-stars that make up the $\theta^1$ or Trapezium cluster have ionized the adjacent gas and carved out a blister or cavity shaped region on the surface of the Orion Molecular Cloud \citep{zuckerman73, genzel89, odell11, odell09}.  
The UV radiation field generated by the Trapezium cluster is fairly well constrained  \citep{ferland12}, with the O7V star $\theta^1$~Ori~C contributing most of the UV photons.

The
southeastern
edge of the blister is a dense ($n~\gtrsim~10^5$~cm$^{-3}$)  PDR called the ``Orion Bar,'' viewed nearly edge on \citep{tielens93, walmsley00, pellegrini09}.   
The \htwo{} emission from its dissociation front is bright and has been well studied.  The first observations of the \htwo{} emission (e.g., \citealt{hayashi85, gatley87}) found intermediate rovibrational level populations that they interpreted as a combination of pure UV and shock excitation in the \htwo{}.  
Later studies by \citet{hippelein89}, \citet{burton90b} and \citet{parmar91} found that the \htwo{} line widths in the bar are narrow, with Local Standard of Rest (LSR) radial velocities matching the ambient molecular cloud, inconsistent with shocks,
which
typically exhibit significant lateral motion (e.g.. such as observed in Orion KL by \citealt{oh16}).
These authors suggested instead that the emission arises from collisionally modified UV excited \htwo{}.  \citet{luhman98} came to the same conclusion from their observations of 16 \htwo{} lines in the Bar.

The  large spatial scale of the Orion Bar suggests that it is supported in a state of quasi-hydrostatic equilibrium by magnetic pressure that counteracts the radiation pressure from the Trapezium stars  \citep{pellegrini09, shaw09}.  Others, such as \citet{parmar91},
 \citet{goicoechea16}, and
 \citet{salgado16}, argue that the Orion Bar is not in
hydrostatic equilibrium but instead represents a slow moving ($< 4$ km s$^{-1}$) magnetohydrodynamic shock front of swept-up molecular material supported by a strong compressed magnetic field.
Observations of the Orion Bar find that complex molecules in the far-IR,
sub-millimeter,
and radio  \citep{danby88, simon97, young00, bartla03, lis03, parise09,  goicoechea11, goicoechea16} trace relatively cool dense gas
($T=50$ to $250$~K, $n_H = 10^5$ to $10^7$~cm$^{-3}$).
Observations of the collisionally excited pure rotation ($v=0$) lines of \htwo{} \citep{parmar91, allers05, shaw09},
ions such as C$^+$
\citep{tielens93, tauber94, wyrowski97},  and excited molecules formed in the presence of rovibrationally excited \htwo{} \citep{nagy13} trace warmer lower density gas 
($T=250$ to $1000$~K, $n_H=10^4$ to $10^5$~cm$^{-3}$).
The emerging consensus is that the Orion Bar PDR does not consist of a single homogenous slab of gas, but instead is composed of cooler dense molecular clumps or filaments embedded in a warmer lower density medium \citep{burton90, parmar91, meixner93, andree-labsch14}.  However, some have argued against the presence of dense clumps near the dissociation front where the \htwo{} emission is strongest \citep{marconi98, allers05}.
All these observations find that the Orion Bar gas is warmer than models predict, suggesting that an extra heating 
mechanism,
not yet identified, is present.  Several candidate heating mechanisms have been proposed including an enhanced flux of cosmic rays trapped by a strong magnetic field \citep{pellegrini07, pellegrini09, shaw09}, a larger than expected number of photoelectrons from grains  \citep{allers05},
X-rays
emitted by young stars in the Orion Nebula \citep{shaw09}, or collisional 
de-excitation
of formation pumped \htwo{} \citep{lebourlot12}.

In this paper, we use \htwo{} to probe the physical conditions and processes in the zone of the Orion Bar dissociation front.
Section \ref{sec:obs} describes our deep near-infrared spectrum of the Orion Bar, taken at the location of the peak \htwo{} surface brightness, with the Immersion Grating INfrared Spectrometer (IGRINS).
In \sectionsymbol{}~\ref{sec:reduction-extraction}, we describe the initial data reduction, wavelength calibration, flux calibration, telluric correction, method for extracting \htwo{} line fluxes, and
effects
of dust extinction.
We discuss how we convert the line fluxes into rovibrational level populations in  \sectionsymbol{}~\ref{sec:analysis}.
Our spectrum contains a larger number of 
\htwo{} rovibrational transition emission lines at higher spectral resolution than all previous near-IR studies of the Orion Bar.
The lines are all observed simultaneously through the same slit and cover a wide range of \htwo{} upper vibrational states from $v = 1$~to~$11$ with level energies up to
50,000~K
above the ground state ($v=0,~J=0$).
This gives us an excellent handle on the excitation mechanisms of the \htwo{}.
In \sectionsymbol{}~\ref{sec:modeling}, we compare the observed \htwo{} rovibrational level populations to those predicted by Cloudy models \citep{shaw05, cloudy13}, to check whether we can match the observed level populations in the Orion Bar, determine which models provide the best match, and discuss the implications of the temperature and density of the \htwo{} emitting gas derived from the model fits.
We present our summary and conclusions in \sectionsymbol{}~\ref{sec:conclusions}.

\section{Observations} \label{sec:obs}

\begin{figure*}
\hspace{-0.5cm}
\includegraphics[width=4.3in]{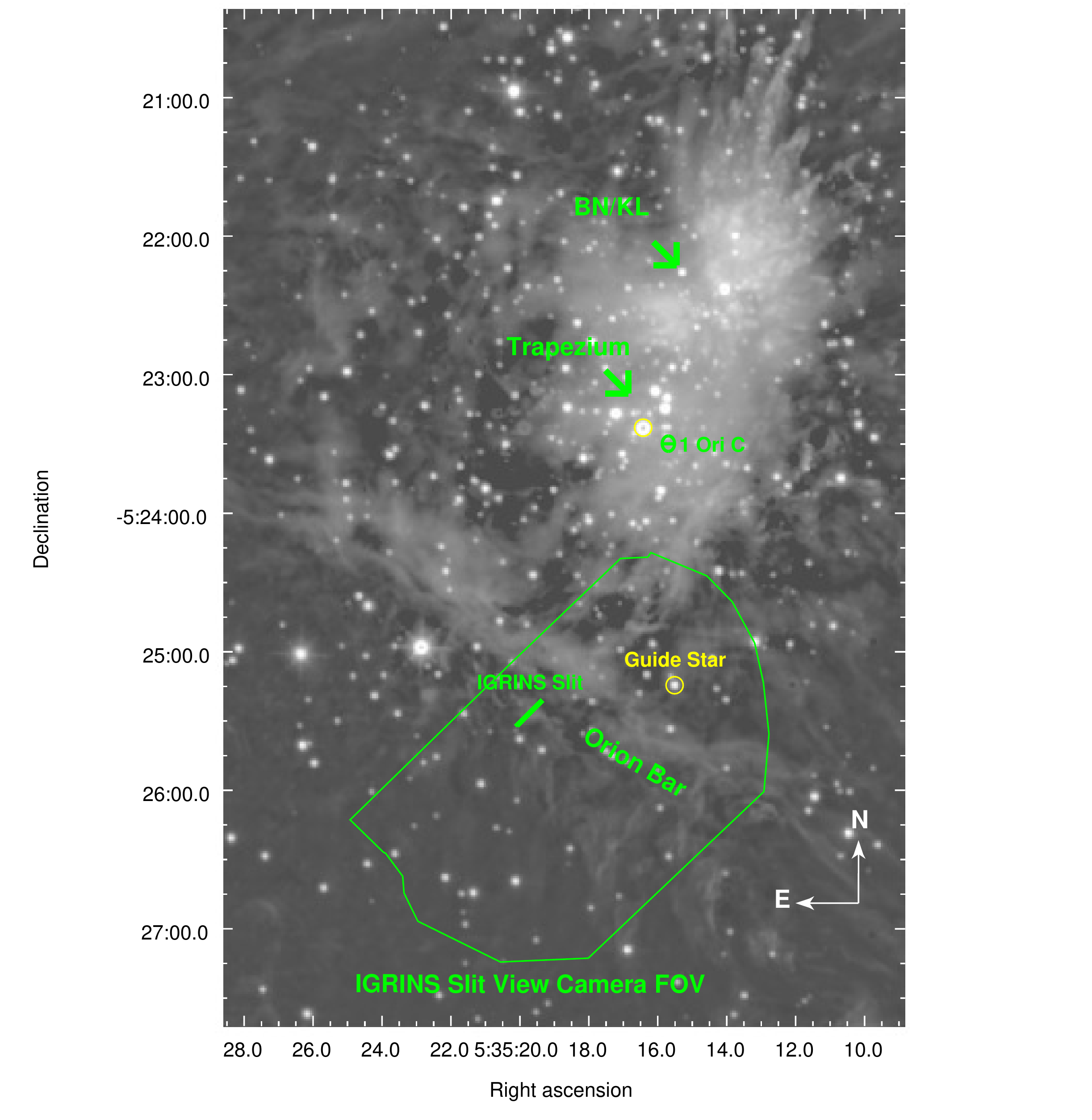}
\hspace{-1.7cm}
\includegraphics[width=3.7in]{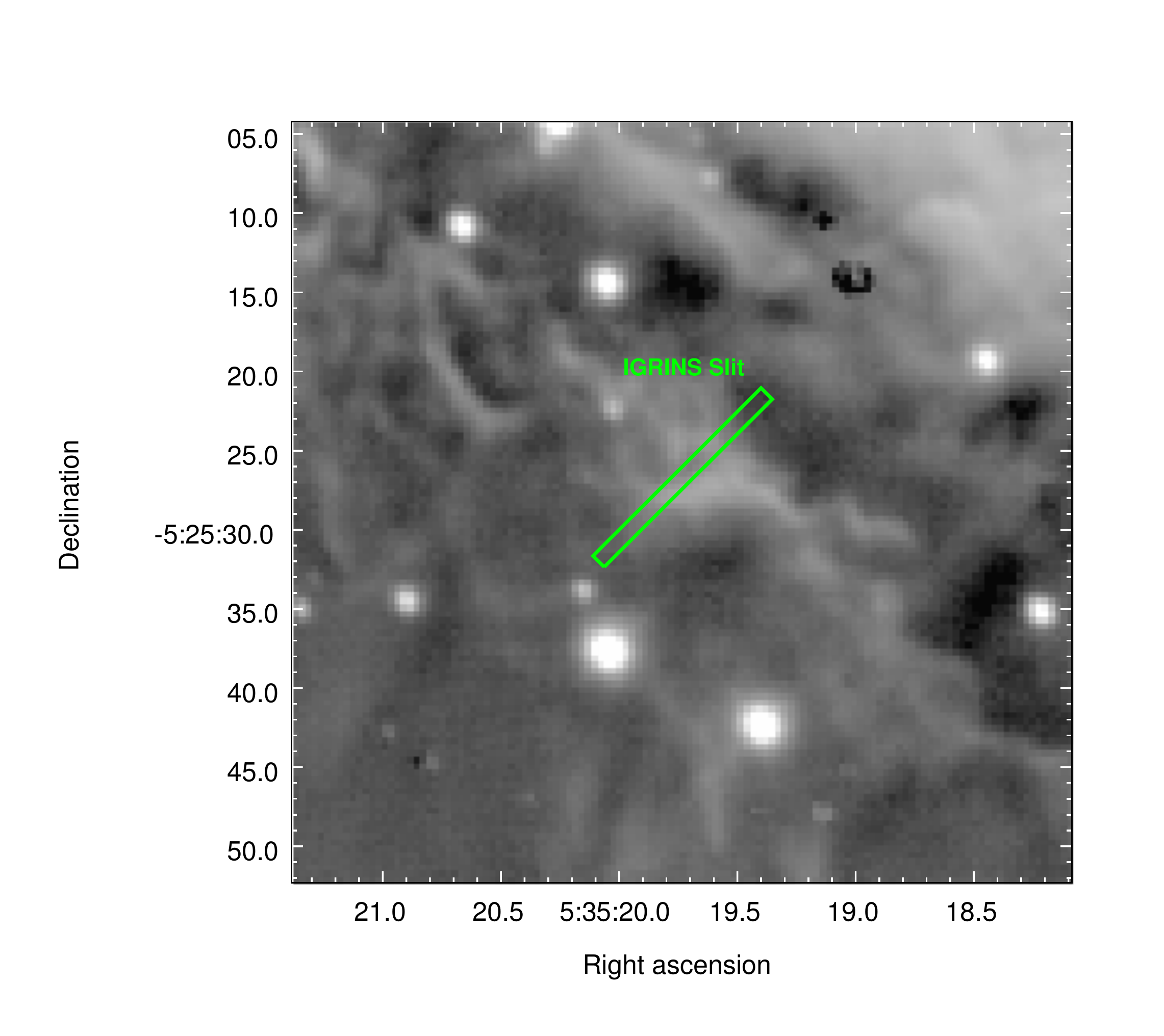}
\caption{
Left: finder
chart showing the location of our pointing on the Orion Bar (slit to scale centered on 
05\hours{}35\minutes{}19\seconds{}73, -05\degrees{}25\arcminutes{}26\arcseconds{}7, J2000),
the guide star V1501 Ori 
(5\hours{}35\minutes{}15\seconds{}55, -05\degrees{}24\arcminutes{}14\arcseconds{}0, J2000),
the FOV of the IGRINS slit-viewing camera, and various features of the Orion Nebula including the Orion Bar, the Trapezium Cluster, Orion BN/KL, and the O-star $\theta^1$ 
C, which
is the primary source of UV photons interacting with the Orion Bar.  IGRINS was rotated to slit PA 135\degrees{} counterclockwise from
the north.
This
narrow-band
image of the \htwo{} 1-0~S(1) line at 2.12183  $\mu$m used for the finder chart is from the Database of Near-IR Narrow-band WFCAM Images for Star Formation hosted by the JAC: \url{http://www.ukirt.hawaii.edu/TAP/singles.html}.  The image was taken with WFCAM on UKIRT by \citet{davis09}.  This
narrow-band
\htwo{} image is not continuum subtracted, and thus it includes scattered starlight.
Right: zoomed-in
view of the slit position.
\\}
\label{fig:full_finder}
\end{figure*}

The data were taken with the
IGRINS on the 2.7 m Harlan J. Smith Telescope at the McDonald Observatory on the night of 2014 October 24 UT.  IGRINS is a near-infrared cross-dispersed echelle spectrometer that uses a silicon immersion grating to achieve
%\textbf{
%a
high
%}
spectral resolution of  $R = \lambda / \Delta \lambda \approx 45000$ or 7.5 km s$^{-1}$ in two separate 
H- and K-band
channels ($1.45-2.45\ \mu$m; \citealt{park14}).
Dark and flat calibration frames were taken with the internal IGRINS calibration unit at the beginning of the night.
IGRINS contains a fixed slit that subtends approximately $1'' \times 15 ''$ on the sky when the instrument is mounted on the 2.7 m telescope at McDonald Observatory.  For the Orion Bar observations, IGRINS was rotated at the Cassegrain focus to set the slit position angle (PA) on the sky  to 135\degrees{} counterclockwise from north, perpendicular to the dissociation front.  Pointing and tracking were performed with the IGRINS slit-viewing camera, which images a  $\sim 2' \times 3'$ field surrounding the slit in the K-band every
10 s.
We used the nearby star V1501~Ori at
\mbox{05\hours{}35\minutes{}15\seconds{}55}, \mbox{-05\degrees{}24\arcminutes{}14\arcseconds{}0}
(J2000) for offslit guiding.  
The center of the slit was positioned at
05\hours{}35\minutes{}19\seconds{}73, -05\degrees{}25\arcminutes{}26\arcseconds{}7
(J2000), within the maps from \citet{allers05}.
Figure  \ref{fig:full_finder}
shows the finder chart and the IGRINS slit position and angle superposed on the Orion Bar.  
We took three 10 minute exposures on
the target
and three 10 minute exposures on
the sky
(30\arcminutes{} south and
30\arcminutes{} west
of the target).
The sky exposures were used to subtract various backgrounds such as telluric OH emission, H$_2$O emission, thermal emission, bias signal, and dark current.
For telluric correction and relative flux calibration, 
we observed the A0V star HD~34317 with four exposures of
60 s
each,
nodding the star between two positions along the slit, around the same
airmass and time
we observed the Orion Bar.
We subtracted one nod position from the other to remove sky and background while preserving the signal at each position.

\section{Data Reduction and Line Flux Extraction} \label{sec:reduction-extraction}

\subsection{Basic Data Reduction and Wavelength Calibration}
%We remove cosmic rays and bad pixels with a custom Python routine before running the data reduction pipeline.
%Cosmic rays are removed from the target and sky frames by subtracting the median of all target and sky frames respectively, and sigma clipping the residuals.
%Hot pixels are removed by summing all dark calibration frames and sigma clipping to find and remove the brightest pixels.  Dead pixels are removed by summing the flat calibration frames, median smoothing, and then subtracting the median smoothed flats from the unsmoothed summed ones and sigma clipping to remove the dead pixels which show up as outliers.

To reduce the data, we run the data reduction pipeline (IGRINS Pipeline Package
[PLP],\footnote{IGRINS Pipeline Package (PLP): \url{https://github.com/igrins/plp}.} \citealt{plp}).
The PLP finds and subtracts the vertical and horizontal medians of the inter-order pixels to remove the readout pattern from each raw frame.
Flat frames are combined to create a master flat, which is used to correct pixel to pixel variations on the detector and to trace the aperture for each order in the echellograms.
The A0V standard star spectrum
is optimally extracted with a weighted sum along the positive and negative traces of the star, which are then summed into a single one-dimensional spectrum.  The spectrum of the Orion Bar, which is spatially extended and fills the whole slit,
is extracted by subtracting the sky frames from the science frames with no weighting.
Cosmic rays are identified and masked by running the Python version of LA-Cosmic \citep{dokkum2001} on the reduced science frames divided the square of their variance.
We use OH emission lines from the sky frames 
as an initial estimate of the wavelength solution by fitting 2D Chebyshev polynomials to the
H- and K-band
echellograms.  The polynomial solution is refined by cross-correlating the telluric absorption lines in the A0V standard star spectrum with predictions from an atmospheric model generated with the Telfit\footnote{Telfit: \url{http://www.as.utexas.edu/~kgulliks/projects.html}.} code by  \citet{guillikson14}.
The final wavelength solution has sub-pixel accuracy with a typical uncertainty of $\pm 0.5$~km~s$^{-1}$ or $< 6\times10^{-6}$~$\mu$m at any given point in the spectrum.  We confirm our final wavelength solution by finding no significant difference between it and  solutions derived using a ThAr arc lamp or the OH sky emission lines.

For the rest of this paper, we carry out our calibrations and analysis using our publicly available ``plotspec''\footnote{Plotspec: \url{https://github.com/kfkaplan/plotspec}.} python code designed for analyzing the reduced 2D IGRINS spectra of emission line nebulae.

\subsection{Telluric Correction and Relative Flux Calibration}
Calculation of relative column densities of \htwo{} rovibrational states requires reliable flux ratios for all observed lines across the full wavelength range covered by IGRINS. 
To obtain a reliable relative spectrophotometric flux calibration,
we need to correct for atmospheric (telluric) absorption lines, atmospheric transmission, instrumental throughput, and detector response.
A0V stars have a
well-known
continuum shape, broad \hnaught{} absorption lines, and weak metal lines, making them preferred standards for telluric correction and relative flux calibration in the near-IR.

We adopt a technique similar to that used for telluric correction and relative flux calibration in the SpeX data reduction package \textit{Spextool}\footnote{Spextool: \url{http://irtfweb.ifa.hawaii.edu/~spex/}.} \citep{vacca2003}.
We assume
that
every A0V star has a continuum shape similar to that of Vega, and modify the model spectrum of Vega {\it vegallpr25.50000resam5} by R.
Kurucz\footnote{R. Kurucz synthetic stellar spectra: \url{http://kurucz.harvard.edu/stars.html}.} to match the spectrum of the observed A0V standard star.
First, we mask out the \hnaught{} absorption lines in the model Vega spectrum and fit a cubic spline to the regions between the absorption lines to define the continuum.  Next, we artificially redden this continuum to match the A0V standard HD 34317 using the near-IR extinction law from \citet{reike85}, with $E(B-V)$ calculated from the standard star's
\textit{B} and \textit{V}
magnitudes.
The \hnaught{} absorption lines in the spectrum of the standard star are fit by scaling and Gaussian smoothing the \hnaught{} lines in the model Vega spectrum and adding them to the artificially reddened continuum to create a synthetic spectrum of the standard star.
This synthetic spectrum represents our estimate of the intrinsic spectrum of the standard star.
Finally, we divide the IGRINS spectrum of the A0V standard HD 34317 by the synthetic spectrum to derive the counts-to-flux ratio at each wavelength, and simultaneously apply  the telluric correction and relative flux calibration by dividing the science spectrum by this ratio.

\subsection{Continuum and Residual OH Removal}

The faint continuum arises from a combination of starlight scattered off dust grains along with free-free and
\mbox{bound-free}
emission from the ionized gas.  The continuum is subtracted from each order using a two step robust median filtering technique.   We start by assuming that the spatial profile of the continuum along the slit has the same shape for all wavelengths in an order, but that the total flux from the continuum may vary with wavelength.  For the first step, we estimate the continuum's spatial profile by finding the median for each row of pixels along the spectral axis.  For the second step, we normalize the estimated spatial profile flux at each wavelength to the median of the surrounding $\pm 187$ pixels.
The normalized median continuum spatial profile across each order is then subtracted from the order.
This technique fits the continuum well, while ignoring narrow features in the
spectrum,
such as emission and absorption lines, bad pixels, or regions with a large amount of telluric absorption. 
After continuum subtraction, we splice all orders together into a single 2D spectrum on a single wavelength grid.

\begin{figure*}
 \centering
 \vspace{-0.5cm}
 \hspace{-0.11in}
 \includegraphics[height=5.5in]{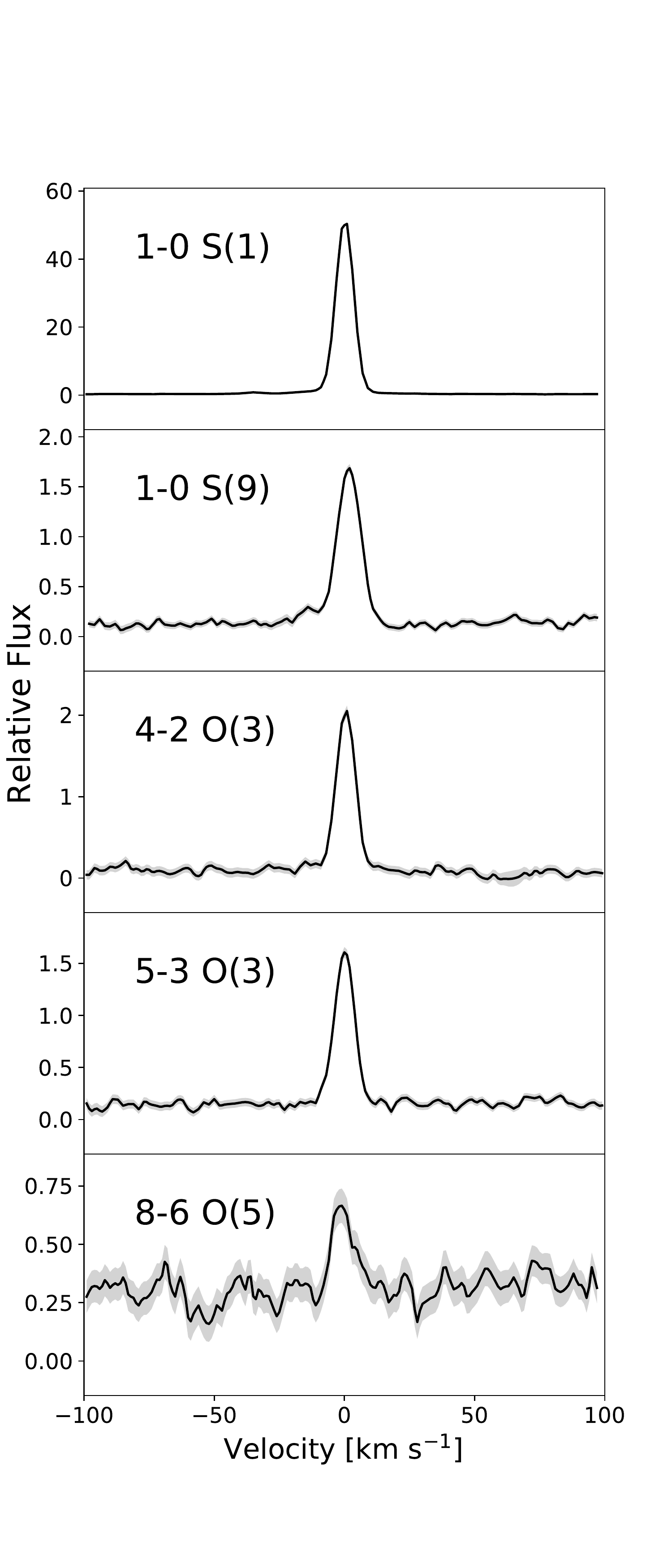}
\includegraphics[height=5.5in]{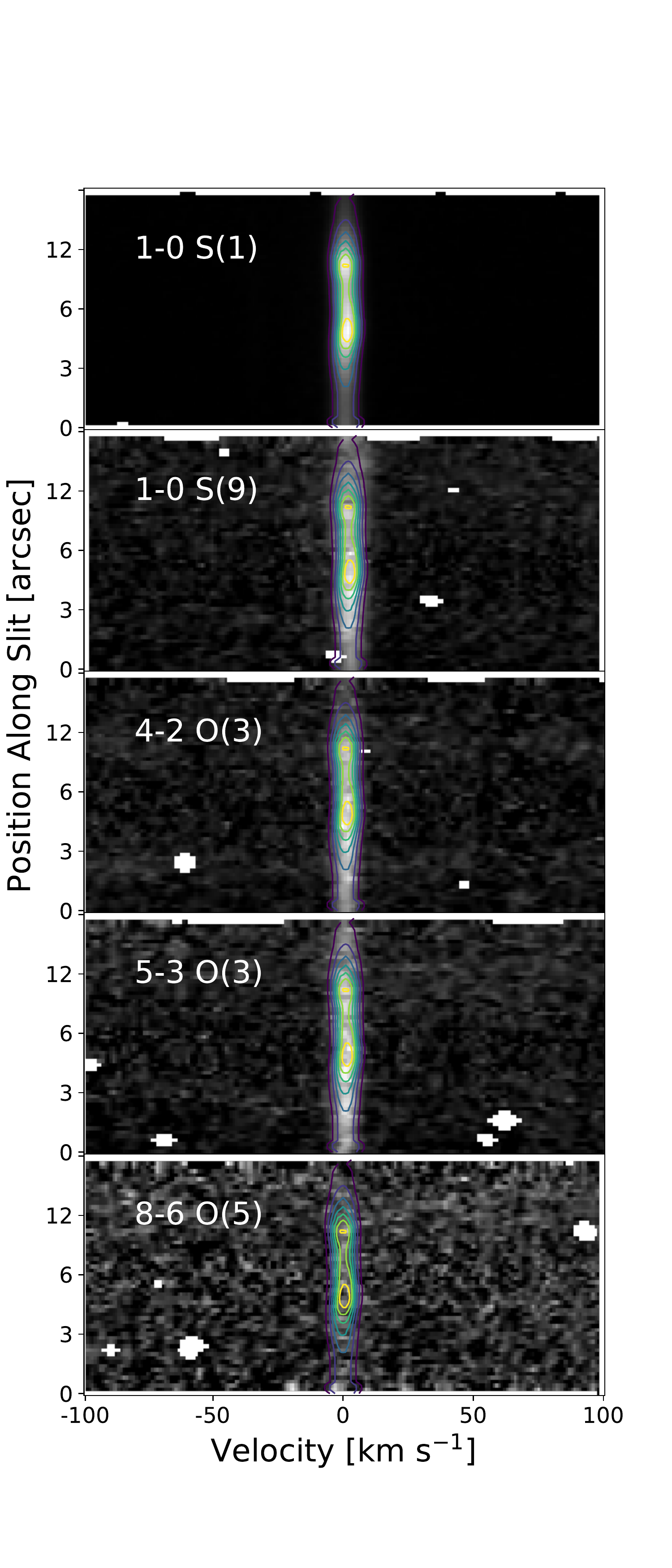}
\vspace{-0.5cm}
\caption{ One-dimensional \htwo{} rovibrational line profiles (left) and two-dimensional PV diagrams (right) for the 1-0~S(1), 1-0~S(9), 4-2~O(3),
5-3~O(3),
and 8-6~O(5) transitions, which arise from a range of upper $v$ and $J$ states.  The dotted lines and light
gray
shading in the 1D line profiles shows the 1$\sigma$ statistical uncertainty.  The 2D color contours show the weights used to extract the flux for each line, as defined in  Equations \ref{eq:weighting} and \ref{eq:weighting-normalized}.
The white spots on the 2D PV diagrams are masked out cosmic rays.
}
\label{fig:line-profiles}
\end{figure*}

Since telluric OH emission lines vary in flux over time, OH line residuals are a possible source of confusion and could in some cases be misidentified as \htwo{} lines.
However, OH residuals can be easily identified using the list of \citet{rousselot2000}.  Furthermore, their fluxes are roughly uniform in the spatial dimension along the 15\arcsec{} slit, whereas the \htwo{} lines vary in brightness across the slit (as
seen, for example,
in Figure \ref{fig:line-profiles}).
To minimize the effect of OH residuals, we apply a first order correction by taking the difference between the first and last sky frames to estimate the variability of the OH lines.  We then scale and subtract the difference in sky frames from the science frames, removing most of the flux from OH residuals.

\subsection{Line Wavelengths} \label{sec:wavelengths}

The spectral resolution of IGRINS is large enough that we can use it to test the rovibrational energy levels used to calculate wavelengths for our line list.
To correct for the net LSR, solar, and barycentric velocities, 
the line vacuum wavelengths were shifted by 8.0~km~s$^{-1}$, derived from fitting the centroid offset of the 1-0~S(1) line.
The vacuum wavelengths for the \htwo{} lines are calculated from the theoretical ground electronic state rovibrational energy levels given in \citet{komasa11}, and reported in column 1 of Table \ref{tab:coldens}.
We measure the difference between the observed line centroids and the theoretical wavelengths (given as $\Delta\lambda$ in column 2 of Table \ref{tab:coldens}).  This difference is well within the wavelength calibration precision of 
$< 6\times10^{-6}$~$\mu$m for most lines, although 1-0~S(6),
2-0 O(9),
 5-3~O(4), 7-5~S(4),
7-5 Q(13),
 8-6~S(3), 9-7~O(4), and 9-7~Q(2) show somewhat larger deviations.
We observe the same differences in wavelengths in other PDRs, confirming that these deviations are real and not caused by kinematics within the Orion Bar or some other issue.  To ensure that all the line centroids are aligned in velocity space, we adjust the wavelengths in our line list to match these small deviations from the theoretical values.

%\afterpage{ %Start table at top of page
\begin{longtable*}{rrrrrrrrrrr}
\caption{\htwo{} lines observed in the Orion Bar}{} \label{tab:coldens} \\
\hline
\multicolumn{1}{c}{$\lambda_{\mbox{\tiny vacuum}}$} & \multicolumn{1}{c}{$\Delta\lambda$} & \multicolumn{1}{c}{\htwo{} line ID} & \multicolumn{1}{c}{$\log_{10} \left(F_i / F_r\right)$} & \multicolumn{1}{c}{S/N} & \multicolumn{1}{c}{$v_u$} & \multicolumn{1}{c}{$J_u$} & \multicolumn{1}{c}{$E_u/k$} & \multicolumn{1}{c}{$\log_{10}\left(A_{ul}\right)$} & \multicolumn{1}{c}{$\ln \left({N_u \over g_u} \big / {N_r \over g_r}\right)$} & \multicolumn{1}{c}{$N_u/N_m$}  \\
\multicolumn{1}{c}{($\mu$m)} & \multicolumn{1}{c}{($10^{-6}$ $\mu$m)} & \multicolumn{1}{c}{} & \multicolumn{1}{c}{} & \multicolumn{1}{c}{} & \multicolumn{1}{c}{} & \multicolumn{1}{c}{} & \multicolumn{1}{c}{(K)} & \multicolumn{1}{c}{[$\log_{10}$(s$^{-1}$)]} & \multicolumn{1}{c}{} & \multicolumn{1}{c}{}\\
\multicolumn{1}{c}{(1)} & \multicolumn{1}{c}{(2)} & \multicolumn{1}{c}{(3)} & \multicolumn{1}{c}{(4)} & \multicolumn{1}{c}{(5)} & \multicolumn{1}{c}{(6)} & \multicolumn{1}{c}{(7)} & \multicolumn{1}{c}{(8)} & \multicolumn{1}{c}{(9)} & \multicolumn{1}{c}{(10)} & \multicolumn{1}{c}{(11)}\\
\hline\hline
\endfirsthead
\hline
\multicolumn{1}{c}{$\lambda_{\mbox{\tiny vacuum}}$} & \multicolumn{1}{c}{$\Delta\lambda$} & \multicolumn{1}{c}{\htwo{} line ID} & \multicolumn{1}{c}{$\log_{10} \left(F_i / F_r\right)$} & \multicolumn{1}{c}{S/N} & \multicolumn{1}{c}{$v_u$} & \multicolumn{1}{c}{$J_u$} & \multicolumn{1}{c}{$E_u/k$} & \multicolumn{1}{c}{$\log_{10}\left(A_{ul}\right)$} & \multicolumn{1}{c}{$\ln \left({N_u \over g_u} \big / {N_r \over g_r}\right)$} & \multicolumn{1}{c}{$N_u/N_m$}  \\
\multicolumn{1}{c}{($\mu$m)} & \multicolumn{1}{c}{($10^{-6}$ $\mu$m)} & \multicolumn{1}{c}{} & \multicolumn{1}{c}{} & \multicolumn{1}{c}{} & \multicolumn{1}{c}{} & \multicolumn{1}{c}{} & \multicolumn{1}{c}{(K)} & \multicolumn{1}{c}{[$\log_{10}$(s$^{-1}$)]} & \multicolumn{1}{c}{} & \multicolumn{1}{c}{}\\
\multicolumn{1}{c}{(1)} & \multicolumn{1}{c}{(2)} & \multicolumn{1}{c}{(3)} & \multicolumn{1}{c}{(4)} & \multicolumn{1}{c}{(5)} & \multicolumn{1}{c}{(6)} & \multicolumn{1}{c}{(7)} & \multicolumn{1}{c}{(8)} & \multicolumn{1}{c}{(9)} & \multicolumn{1}{c}{(10)} & \multicolumn{1}{c}{(11)}\\
\hline\hline
\endhead
\hline
\endfoot
\hline
\multicolumn{11}{l}{\parbox{0.9\textwidth}{Columns are as follows. (1) The \htwo{} line vacuum wavelength in $\mu$m calculated from the ground electronic state rovibrational energy levels in \citet{komasa11}.  See \sectionsymbol{}~\ref{sec:wavelengths} for more details. (2) The observed line centroid wavelength (in the Orion Bar rest frame) minus the expected theoretical line wavelength calculated from the level energies in \citet{komasa11} in units of $10^{-6} \mu$m (\sectionsymbol{}~\ref{sec:wavelengths}). (3) \htwo{} line rovibrational identifications in spectroscopic notation in the format ``W-X Y(Z).''  W and X denote the transition's upper and lower $v$ states.  Y denotes the change in $J$, where S is $\Delta J = -2$, Q is  $\Delta J = 0$, and O is $\Delta J = +2$.   Z denotes the upper $J$ state. (4) The base 10 logarithm of the line flux $F_i$ normalized to the 4-2 O(3) reference line flux $F_r$ (\sectionsymbol{}~\ref{sec:extract}).  (5) The signal-to-noise ratio for the line flux (\sectionsymbol{}~\ref{sec:extract}). (6) The transition's upper vibrational state. (7) The transition's upper rotational state. (8) The energy of the upper state $E_u$ above the ground  ($v=0$, $J=0$) divided by the Boltzmann constant $k$ to convert the energies into temperature units (\sectionsymbol{}~\ref{sec:h2-level-pop}).  (9) The  base 10 logarithm of the rovibrational radiative transition probability $A_{ul}$ from \citet{wolniewicz98}, in units of $s^{-1}$  (\sectionsymbol{}~\ref{sec:h2-pop-interp}).  (10) The natural logarithm of the column density in a transition's upper state $N_u$ divided by the quantum degeneracy $g_u$, normalized to $N_r /g_r$ for the reference line 4-2 O(3) (\sectionsymbol{}~\ref{sec:h2-pop-interp} and  \sectionsymbol{}~\ref{sec:h2-level-pop}).  This is the value plotted in the excitation diagram shown in Figure \ref{fig:model-best}.  (11) The ratio of the observed column density of the transition's upper state $N_u$ to the column density predicted by our best-fit model $N_m$ (\sectionsymbol{}~\ref{sec:const-models}), as shown in the bottom of Figure \ref{fig:model-best}.}}
\endlastfoot
2.406592 & 0.95 &  1-0 Q(1) & $1.273^{+0.001}_{-0.001} $ & 817.4 & 1 & 1 &  6149 & -6.37 & $3.982^{+0.001}_{-0.001}$  & 0.60 \\
2.413439 & 1.67 &  1-0 Q(2) & $0.762^{+0.001}_{-0.001} $ & 399.7 & 1 & 2 &  6471 & -6.52 & $3.743^{+0.002}_{-0.003}$  & 0.81 \\
2.223290 & 0.72 &  1-0 S(0) & $0.731^{+0.001}_{-0.001} $ & 580.0 & 1 & 2 &  6471 & -6.60 & $3.771^{+0.002}_{-0.002}$  & 0.83 \\
2.423730 & 5.72 &  1-0 Q(3) & $1.022^{+0.001}_{-0.001} $ & 488.6 & 1 & 3 &  6951 & -6.56 & $2.996^{+0.002}_{-0.002}$  & 0.86 \\
2.121834 & 0.00 &  1-0 S(1) & $1.193^{+0.000}_{-0.000} $ & 1031.9 & 1 & 3 &  6951 & -6.46 & $3.035^{+0.001}_{-0.001}$  & 0.90 \\
2.437489 & 0.00 &  1-0 Q(4) & $0.417^{+0.002}_{-0.002} $ & 225.6 & 1 & 4 &  7584 & -6.58 & $2.504^{+0.004}_{-0.004}$  & 1.36 \\
2.033758 & -0.72 &  1-0 S(2) & $0.643^{+0.001}_{-0.001} $ & 422.1 & 1 & 4 &  7584 & -6.40 & $2.436^{+0.002}_{-0.002}$  & 1.27 \\
1.957559 & -2.62 &  1-0 S(3) & $0.959^{+0.001}_{-0.001} $ & 339.8 & 1 & 5 &  8365 & -6.38 & $1.772^{+0.003}_{-0.003}$  & 1.73 \\
1.788050 & -21.46 &  1-0 S(6) & $-0.091^{+0.005}_{-0.006} $ & 78.8 & 1 & 8 & 11521 & -6.45 & $0.099^{+0.013}_{-0.013}$  & 0.60 \\
1.747955 & -2.86 &  1-0 S(7) & $0.285^{+0.003}_{-0.003} $ & 132.7 & 1 & 9 & 12817 & -6.53 & $-0.096^{+0.008}_{-0.008}$  & 0.67 \\
1.714738 & -2.26 &  1-0 S(8) & $-0.435^{+0.015}_{-0.015} $ & 29.2 & 1 & 10 & 14220 & -6.63 & $-0.532^{+0.034}_{-0.035}$  & 0.40 \\
1.687761 & -3.93 &  1-0 S(9) & $-0.202^{+0.005}_{-0.005} $ & 89.3 & 1 & 11 & 15721 & -6.78 & $-0.869^{+0.011}_{-0.011}$  & 0.43 \\
1.666475 & -1.67 &  1-0 S(10) & $-1.029^{+0.031}_{-0.034} $ & 13.4 & 1 & 12 & 17311 & -6.98 & $-1.306^{+0.072}_{-0.078}$  & 0.32 \\
1.650413 & 0.60 &  1-0 S(11) & $-1.021^{+0.027}_{-0.029} $ & 15.4 & 1 & 13 & 18979 & -7.27 & $-1.789^{+0.063}_{-0.067}$  & 0.30 \\
2.355605 & -2.38 &  2-1 S(0) & $0.000^{+0.004}_{-0.004} $ & 97.8 & 2 & 2 & 12095 & -6.43 & $1.769^{+0.010}_{-0.010}$  & 1.92 \\
2.247716 & 1.67 &  2-1 S(1) & $0.465^{+0.001}_{-0.001} $ & 376.5 & 2 & 3 & 12550 & -6.30 & $1.057^{+0.003}_{-0.003}$  & 1.71 \\
2.154216 & -1.43 &  2-1 S(2) & $-0.012^{+0.003}_{-0.003} $ & 159.3 & 2 & 4 & 13150 & -6.25 & $0.645^{+0.006}_{-0.006}$  & 1.30 \\
2.073482 & 0.00 &  2-1 S(3) & $0.408^{+0.001}_{-0.001} $ & 323.5 & 2 & 5 & 13890 & -6.24 & $0.244^{+0.003}_{-0.003}$  & 1.73 \\
1.679641 & 9.66 &  2-0 O(9) & $-1.421^{+0.060}_{-0.069} $ & 6.8 & 2 & 7 & 15763 & -7.89 & $-0.692^{+0.137}_{-0.159}$  & 1.18 \\
1.522033 & -0.60 &  3-1 O(5) & $-0.235^{+0.009}_{-0.009} $ & 46.6 & 3 & 3 & 17818 & -6.70 & $-0.025^{+0.021}_{-0.022}$  & 1.18 \\
2.386471 & -3.10 &  3-2 S(1) & $-0.006^{+0.006}_{-0.006} $ & 68.0 & 3 & 3 & 17818 & -6.29 & $-0.004^{+0.015}_{-0.015}$  & 1.21 \\
1.581171 & 1.55 &  3-1 O(6) & $-0.837^{+0.027}_{-0.029} $ & 15.6 & 3 & 4 & 18386 & -6.86 & $-0.163^{+0.062}_{-0.066}$  & 1.18 \\
2.287045 & 0.72 &  3-2 S(2) & $-0.392^{+0.006}_{-0.006} $ & 72.9 & 3 & 4 & 18386 & -6.25 & $-0.181^{+0.014}_{-0.014}$  & 1.16 \\
2.201399 & 0.72 &  3-2 S(3) & $-0.013^{+0.003}_{-0.003} $ & 150.1 & 3 & 5 & 19086 & -6.25 & $-0.645^{+0.007}_{-0.007}$  & 1.55 \\
2.128015 & 1.43 &  3-2 S(4) & $-0.508^{+0.006}_{-0.006} $ & 69.3 & 3 & 6 & 19911 & -6.28 & $-0.813^{+0.014}_{-0.015}$  & 1.49 \\
2.065584 & 1.43 &  3-2 S(5) & $-0.181^{+0.004}_{-0.004} $ & 100.1 & 3 & 7 & 20856 & -6.34 & $-1.182^{+0.010}_{-0.010}$  & 1.95 \\
1.509865 & 0.00 &  4-2 O(3) & $0.000^{+0.006}_{-0.006} $ & 70.1 & 4 & 1 & 22079 & -6.11 & $0.000^{+0.014}_{-0.014}$  & 1.00 \\
1.563515 & 1.07 &  4-2 O(4) & $-0.489^{+0.010}_{-0.010} $ & 43.8 & 4 & 2 & 22352 & -6.29 & $-0.107^{+0.023}_{-0.023}$  & 1.00 \\
1.622299 & -2.74 &  4-2 O(5) & $-0.328^{+0.009}_{-0.009} $ & 47.4 & 4 & 3 & 22759 & -6.44 & $-0.791^{+0.021}_{-0.021}$  & 0.92 \\
1.686462 & 2.26 &  4-2 O(6) & $-1.045^{+0.033}_{-0.036} $ & 12.7 & 4 & 4 & 23295 & -6.58 & $-1.224^{+0.076}_{-0.082}$  & 0.71 \\
1.756281 & 0.00 &  4-2 O(7) & $-0.754^{+0.014}_{-0.014} $ & 30.6 & 4 & 5 & 23955 & -6.73 & $-1.475^{+0.032}_{-0.033}$  & 1.20 \\
2.266764 & 2.15 &  4-3 S(4) & $-1.007^{+0.040}_{-0.044} $ & 10.5 & 4 & 6 & 24733 & -6.39 & $-1.637^{+0.091}_{-0.100}$  & 1.07 \\
2.200974 & 0.72 &  4-3 S(5) & $-0.690^{+0.012}_{-0.012} $ & 35.7 & 4 & 7 & 25623 & -6.49 & $-1.957^{+0.028}_{-0.028}$  & 1.59 \\
2.145873 & -0.72 &  4-3 S(6) & $-1.322^{+0.042}_{-0.047} $ & 9.8 & 4 & 8 & 26616 & -6.64 & $-2.124^{+0.097}_{-0.108}$  & 1.03 \\
2.099586 & 5.48 &  4-2 O(11) & $-1.647^{+0.084}_{-0.104} $ & 4.7 & 4 & 9 & 27706 & -7.36 & $-2.449^{+0.193}_{-0.239}$  & 1.45 \\
2.100426 & 4.29 &  4-3 S(7) & $-1.192^{+0.030}_{-0.033} $ & 13.8 & 4 & 9 & 27706 & -6.86 & $-2.542^{+0.070}_{-0.075}$  & 1.33 \\
1.549455 & -2.62 &  4-2 Q(11) & $-0.819^{+0.022}_{-0.023} $ & 19.6 & 4 & 11 & 30139 & -6.34 & $-3.372^{+0.050}_{-0.052}$  & 0.72 \\
1.560736 & -1.55 &  5-3 O(2) & $-0.509^{+0.012}_{-0.012} $ & 35.8 & 5 & 0 & 26606 & -5.65 & $-0.020^{+0.028}_{-0.028}$  & 1.30 \\
1.613520 & 1.55 &  5-3 O(3) & $-0.211^{+0.007}_{-0.007} $ & 64.9 & 5 & 1 & 26735 & -5.95 & $-0.787^{+0.015}_{-0.016}$  & 0.76 \\
1.671814 & 8.94 &  5-3 O(4) & $-0.671^{+0.016}_{-0.016} $ & 27.3 & 5 & 2 & 26992 & -6.12 & $-0.846^{+0.036}_{-0.037}$  & 0.77 \\
1.515792 & -5.01 &  5-3 Q(4) & $-0.614^{+0.021}_{-0.022} $ & 20.1 & 5 & 4 & 27878 & -6.13 & $-1.367^{+0.049}_{-0.051}$  & 0.98 \\
1.528648 & 1.07 &  5-3 Q(5) & $-0.320^{+0.013}_{-0.013} $ & 34.1 & 5 & 5 & 28498 & -6.14 & $-1.954^{+0.029}_{-0.030}$  & 1.30 \\
2.057127 & -4.77 &  5-3 O(9) & $-1.313^{+0.055}_{-0.063} $ & 7.4 & 5 & 7 & 30063 & -6.81 & $-2.727^{+0.127}_{-0.146}$  & 1.27 \\
1.562635 & -2.03 &  5-3 Q(7) & $-0.552^{+0.012}_{-0.012} $ & 36.9 & 5 & 7 & 30063 & -6.17 & $-2.728^{+0.027}_{-0.027}$  & 1.27 \\
1.608398 & -5.36 &  5-3 Q(9) & $-0.628^{+0.015}_{-0.015} $ & 29.3 & 5 & 9 & 32014 & -6.19 & $-3.053^{+0.034}_{-0.035}$  & 1.46 \\
1.675032 & 1.55 &  6-4 O(2) & $-0.747^{+0.014}_{-0.015} $ & 29.6 & 6 & 0 & 30942 & -5.55 & $-0.709^{+0.033}_{-0.034}$  & 1.04 \\
1.601534 & -4.29 &  6-4 Q(1) & $-0.348^{+0.009}_{-0.009} $ & 49.0 & 6 & 1 & 31063 & -5.85 & $-1.339^{+0.020}_{-0.021}$  & 0.74 \\
1.732641 & -0.60 &  6-4 O(3) & $-0.388^{+0.012}_{-0.012} $ & 37.0 & 6 & 1 & 31063 & -5.85 & $-1.354^{+0.027}_{-0.027}$  & 0.73 \\
1.536891 & 0.60 &  6-4 S(0) & $-0.706^{+0.018}_{-0.019} $ & 23.4 & 6 & 2 & 31303 & -6.08 & $-1.091^{+0.042}_{-0.044}$  & 0.90 \\
1.607390 & -1.67 &  6-4 Q(2) & $-0.665^{+0.014}_{-0.015} $ & 29.6 & 6 & 2 & 31303 & -6.00 & $-1.134^{+0.033}_{-0.034}$  & 0.86 \\
1.796524 & 0.60 &  6-4 O(4) & $-0.734^{+0.023}_{-0.025} $ & 18.2 & 6 & 2 & 31303 & -6.01 & $-1.169^{+0.054}_{-0.057}$  & 0.83 \\
1.501560 & 2.03 &  6-4 S(1) & $-0.270^{+0.010}_{-0.011} $ & 41.3 & 6 & 3 & 31661 & -5.94 & $-1.872^{+0.024}_{-0.024}$  & 0.95 \\
1.616224 & -1.67 &  6-4 Q(3) & $-0.468^{+0.008}_{-0.009} $ & 50.7 & 6 & 3 & 31661 & -6.04 & $-2.031^{+0.020}_{-0.020}$  & 0.81 \\
1.628094 & -0.60 &  6-4 Q(4) & $-0.929^{+0.025}_{-0.026} $ & 17.1 & 6 & 4 & 32132 & -6.06 & $-2.197^{+0.057}_{-0.060}$  & 0.73 \\
2.029684 & -3.34 &  6-4 O(7) & $-0.998^{+0.022}_{-0.023} $ & 19.4 & 6 & 5 & 32711 & -6.39 & $-2.659^{+0.050}_{-0.053}$  & 1.02 \\
1.661304 & 4.41 &  6-4 Q(6) & $-1.105^{+0.029}_{-0.031} $ & 14.3 & 6 & 6 & 33394 & -6.08 & $-2.889^{+0.067}_{-0.072}$  & 0.85 \\
1.708041 & 4.53 &  6-4 Q(8) & $-1.244^{+0.040}_{-0.044} $ & 10.4 & 6 & 8 & 35040 & -6.11 & $-3.383^{+0.091}_{-0.101}$  & 0.87 \\
1.728799 & 0.00 &  7-5 Q(1) & $-0.506^{+0.013}_{-0.013} $ & 33.6 & 7 & 1 & 35057 & -5.82 & $-1.702^{+0.029}_{-0.030}$  & 0.78 \\
1.735762 & 4.17 &  7-5 Q(2) & $-1.032^{+0.030}_{-0.032} $ & 14.1 & 7 & 2 & 35281 & -5.97 & $-1.975^{+0.069}_{-0.074}$  & 0.51 \\
1.746280 & -1.79 &  7-5 Q(3) & $-0.670^{+0.020}_{-0.021} $ & 21.7 & 7 & 3 & 35613 & -6.01 & $-2.488^{+0.045}_{-0.047}$  & 0.77 \\
1.620548 & 1.55 &  7-5 S(1) & $-0.539^{+0.010}_{-0.010} $ & 44.8 & 7 & 3 & 35613 & -5.93 & $-2.438^{+0.022}_{-0.023}$  & 0.81 \\
1.760446 & -4.05 &  7-5 Q(4) & $-1.097^{+0.034}_{-0.037} $ & 12.3 & 7 & 4 & 36051 & -6.03 & $-2.571^{+0.078}_{-0.085}$  & 0.82 \\
2.204989 & -0.72 &  7-5 O(7) & $-1.182^{+0.033}_{-0.036} $ & 12.7 & 7 & 5 & 36588 & -6.31 & $-3.188^{+0.076}_{-0.082}$  & 0.87 \\
1.561510 & -1.55 &  7-5 S(3) & $-0.607^{+0.014}_{-0.014} $ & 30.8 & 7 & 5 & 36588 & -5.86 & $-3.250^{+0.032}_{-0.033}$  & 0.82 \\
1.540006 & -6.68 &  7-5 S(4) & $-1.069^{+0.043}_{-0.048} $ & 9.6 & 7 & 6 & 37220 & -5.87 & $-3.382^{+0.100}_{-0.111}$  & 0.72 \\
1.523623 & -0.95 &  7-5 S(5) & $-0.717^{+0.023}_{-0.024} $ & 18.7 & 7 & 7 & 37941 & -5.89 & $-3.758^{+0.052}_{-0.055}$  & 1.05 \\
1.512240 & 2.50 &  7-5 S(6) & $-1.251^{+0.073}_{-0.088} $ & 5.4 & 7 & 8 & 38743 & -5.95 & $-3.902^{+0.169}_{-0.203}$  & 0.74 \\
1.979270 & -1.31 &  7-5 Q(11) & $-1.361^{+0.058}_{-0.066} $ & 7.0 & 7 & 11 & 41558 & -6.18 & $-4.739^{+0.133}_{-0.153}$  & 0.92 \\
2.092904 & 7.63 &  7-5 Q(13) & $-1.576^{+0.059}_{-0.068} $ & 6.9 & 7 & 13 & 43693 & -6.26 & $-5.154^{+0.136}_{-0.157}$  & 0.85 \\
2.041830 & 2.62 &  8-6 O(3) & $-0.684^{+0.012}_{-0.012} $ & 35.3 & 8 & 1 & 38708 & -5.80 & $-2.005^{+0.028}_{-0.029}$  & 0.83 \\
2.210763 & 2.15 &  8-6 O(5) & $-0.993^{+0.034}_{-0.037} $ & 12.1 & 8 & 3 & 39219 & -6.05 & $-2.888^{+0.079}_{-0.086}$  & 0.56 \\
1.763952 & -4.05 &  8-6 S(1) & $-0.864^{+0.040}_{-0.044} $ & 10.4 & 8 & 3 & 39219 & -5.97 & $-3.003^{+0.092}_{-0.101}$  & 0.50 \\
2.310167 & 1.67 &  8-6 O(6) & $-1.529^{+0.091}_{-0.116} $ & 4.3 & 8 & 4 & 39622 & -6.17 & $-2.971^{+0.210}_{-0.266}$  & 0.54 \\
1.701803 & -6.79 &  8-6 S(3) & $-1.014^{+0.031}_{-0.033} $ & 13.5 & 8 & 5 & 40116 & -5.93 & $-3.935^{+0.072}_{-0.077}$  & 0.52 \\
1.664584 & 2.15 &  8-6 S(5) & $-1.169^{+0.035}_{-0.038} $ & 11.9 & 8 & 7 & 41355 & -6.01 & $-4.441^{+0.081}_{-0.088}$  & 0.76 \\
2.172715 & -1.43 &  9-7 O(2) & $-1.280^{+0.042}_{-0.046} $ & 9.9 & 9 & 0 & 41903 & -5.57 & $-1.645^{+0.096}_{-0.106}$  & 0.82 \\
2.073187 & -1.43 &  9-7 Q(1) & $-1.088^{+0.036}_{-0.039} $ & 11.6 & 9 & 1 & 41997 & -5.91 & $-2.661^{+0.082}_{-0.090}$  & 0.93 \\
2.253724 & 1.67 &  9-7 O(3) & $-0.969^{+0.027}_{-0.029} $ & 15.4 & 9 & 1 & 41997 & -5.85 & $-2.451^{+0.063}_{-0.067}$  & 1.15 \\
2.345581 & -8.58 &  9-7 O(4) & $-1.406^{+0.069}_{-0.082} $ & 5.8 & 9 & 2 & 42185 & -5.98 & $-2.518^{+0.159}_{-0.188}$  & 0.51 \\
1.987350 & -4.05 &  9-7 S(0) & $-1.401^{+0.052}_{-0.059} $ & 7.9 & 9 & 2 & 42185 & -6.18 & $-2.204^{+0.119}_{-0.135}$  & 0.70 \\
2.084098 & 9.06 &  9-7 Q(2) & $-1.258^{+0.065}_{-0.076} $ & 6.2 & 9 & 2 & 42185 & -6.06 & $-2.103^{+0.149}_{-0.175}$  & 0.77 \\
2.100664 & 3.58 &  9-7 Q(3) & $-1.237^{+0.031}_{-0.033} $ & 13.7 & 9 & 3 & 42462 & -6.11 & $-3.385^{+0.071}_{-0.076}$  & 0.60 \\
2.151876 & 3.58 &  9-7 Q(5) & $-1.362^{+0.038}_{-0.042} $ & 10.8 & 9 & 5 & 43274 & -6.16 & $-3.971^{+0.088}_{-0.097}$  & 0.69 \\
2.230268 & -4.53 &  9-7 Q(7) & $-1.558^{+0.073}_{-0.088} $ & 5.5 & 9 & 7 & 44392 & -6.23 & $-4.545^{+0.168}_{-0.203}$  & 0.90 \\
1.548849 & 0.60 &  10-7 O(3) & $-1.054^{+0.050}_{-0.056} $ & 8.2 & 10 & 1 & 44903 & -5.98 & $-2.722^{+0.115}_{-0.130}$  & 0.79 \\
2.176855 & 1.67 &  10-8 S(1) & $-1.388^{+0.042}_{-0.046} $ & 9.9 & 10 & 3 & 45317 & -6.27 & $-3.314^{+0.096}_{-0.106}$  & 0.93 \\
1.648305 & -1.67 &  10-7 O(5) & $-1.273^{+0.044}_{-0.049} $ & 9.3 & 10 & 3 & 45317 & -6.29 & $-3.282^{+0.102}_{-0.114}$  & 0.97 \\
\end{longtable*}

%}

Wavelengths calculated from the theoretical rovibrational energy levels of \citet{komasa11} provide much improved agreement between the observed and theoretical line wavelengths over previous values (observed and theoretically derived).
For example, wavelengths calculated from the commonly cited rovibrational energy levels in \citet{dabrowski1984} differ from the observed
line centroids
by up to $10^{-4}$~$\mu$m which is well in excess of the precision of the IGRINS wavelength calibration and differs by 5-10 times more than the wavelengths calculated from \citet{komasa11}.%10~km~s$^{-1}$,

\subsection{Line Flux Extraction}
\label{sec:extract}
We extract line fluxes by interpolating all the \htwo{} rovibrational transition lines we observe in the Orion Bar with IGRINS onto a common position-velocity (PV) grid, on which we use a S/N weighted sum to calculate the flux for each line.
PV maps for each line are created by linearly interpolating from wavelength space to a $\pm 100$ km s$^{-1}$ velocity grid of 1~km~s$^{-1}$ wide pixels.
 \htwo{} lines blended with other lines are removed from consideration.  
Figure \ref{fig:line-profiles} compares the PV diagrams of several of the lines we observed and illustrates our procedure for extracting the flux of each line.  

To extract the line fluxes, we use a flux density weighting scheme designed to scale with  S/N across each line profile. 
Figure \ref{fig:line-profiles} shows that the lines have similar profiles.
We confirm that this is the case for all
the \htwo{} lines, 
by stacking multiple dim lines and comparing the stacked profile to the brightest observed \htwo{} line, 1-0~S(1).
We therefore use the bright 1-0~S(1) line as the basis for our weighting scheme, and calculate the weights $w_{x,v}$ by squaring the flux $F_{x,v}^{\rm 1-0\ S(1)}$ found in each pixel in position ($x$) and velocity ($v$) space for the 1-0~S(1) line:
\begin{equation} \label{eq:weighting}
w_{x,v} = {(F_{x,v}^{\rm 1-0\ S(1)})^2}.
\end{equation}
The weights are then normalized as 
follows:
\begin{equation} \label{eq:weighting-normalized}
w_{x,v} = w_{x,v} / \sum_x \sum_v w_{x,v}.
\end{equation}
The background $B$ per pixel is determined from the median value of all pixels in the PV diagram that
are $\leqslant 0.8\%$ the
flux of the brightest pixel.  The $0.8\%$ limit was chosen to ensure that no line flux ends up in the background determination.
We subtract the background from the flux in each pixel $F_{x,v}$, multiply by the weights $w_{x,v}$, and then sum the result to get the extracted flux $F$:
\begin{equation}
F = \sum_x \sum_v (w_{x,v} (F_{x,v} - B)).
\end{equation}
Each line extraction is visually inspected to ensure that it is a real feature.  Lines that appear to be contaminated by blends, misidentifications such as OH residuals, or noise spikes are rejected. 
For propagation of the statistical uncertainties, the interpolation and extraction process is repeated for the variance reported by the PLP.  Table \ref{tab:coldens} gives the fluxes for all lines with
S/N $> 3$.

\section{Analysis} \label{sec:analysis}

\subsection{Effects of Dust Extinction}

The dense molecular gas of the Orion Bar co-exists with copious amounts of dust.   If there is enough dust in the foreground of the observed \htwo{} emission, the differential extinction across the H and K bands could be significant enough to affect the line ratios we use to derive the rovibrational level populations.
An effective way to measure extinction is to compare
the observed to theoretical
line flux ratios from pairs of lines arising from the same upper level that are widely separated in wavelength.
Two such line pairs in our data with sufficient S/N and widely separated in wavelength are the \mbox{3-1~O(5)/3-2~S(1)} transitions spanning
\mbox{$\lambda = 1.55220 - 2.238645~\mu$m} and the
\mbox{3-1~O(6)/3-2~S(2)}
transitions spanning \mbox{$\lambda = 1.58115 - 2.28703~\mu$m.} 
Assuming the near-IR extinction law from \citet{reike85}, the observed  \mbox{3-1~O(5)/3-2~S(1)} and \mbox{3-1~O(6)/3-2~S(2)} line ratios give extinctions of $A_V = 8.50$ and $8.00$ mag respectively (or $A_K = 0.99$ and $0.93$ mag).  We therefore apply an extinction correction of $A_V = 8.25$ or $A_K = 0.96$ to our spectrum before extracting line fluxes.
This value of the extinction is consistent with the foreground extinction of $A_V \sim 1.3$~mag or  $A_K \sim 0.15$~mag towards the ionized gas \citep{weilbacher15},  allowing for additional extinction between the ionized gas and the region of excited \htwo{}.
%^To test for extinction, we combine the statistical uncertainty in the line ratios with an assumed 10\% systematic uncertainty in the relative flux calibration and telluric correction.
%The observed line ratios for these two pairs of lines differ by $< 1 \sigma$ from their intrinsic line ratios, showing no significant extinction.
%We calculate the $1\sigma$ upper limit for the extinction to be  $A_V < 3.0$ mag or  $A_K < 0.35$ mag by artificially reddening the Orion Bar spectrum with the near-IR extinction law from \citet{reike85}.
%This is consistent with the foreground extinction towards the ionized gas at the slit location of $A_V \sim 1.3$~mag or  $A_K \sim 0.15$~mag  measured from the Balmer decrement by \citet{weilbacher15}, although this measurement does not take into account the Bar's internal extinction.
Our value for extinction in the Bar is lower than the values of $A_K = 2.3 \pm 0.8$~mag and $2.6 \pm 0.7$~mag  for two regions in the Bar $\sim 22$\arcsec{} NE of the slit measured by \citet{luhman98}.    
However, it is possible that the internal extinction is variable depending on the chosen sightline, and that the bright \htwo{} emitting region we targeted is a sightline with low internal extinction.

\subsection{Calculating \htwo{} Level Populations}\label{sec:h2-pop-interp}

The near-IR \htwo{} lines are optically thin and the line fluxes are linearly proportional to the column density of molecules in the upper states of the transitions.
We calculate the column density of \htwo{} in the upper state $N_{u}$ from the following
equation,
\begin{equation}
N_u = {F_{ul} \over \Delta E_{ul} h c A_{ul}},
\end{equation}
where $F_{ul}$ is the flux of the radiative transition from upper ($u$) to lower ($l$) rovibrational states, $\Delta E_{ul}$ is the difference in energy between the states in wavenumbers (cm$^{-1}$), $A_{ul}$ is the transition probability (s$^{-1}$) (we use the values from
\citealt{wolniewicz98}, which
are the same ones used in Cloudy), $h$ is Planck's constant, and $c$ is the speed of light. 

In the Orion Bar, we measure relative fluxes for
87
lines with S/N $> 3$, yielding the relative $N_u$ values reported in Table \ref{tab:coldens}. 
These values are normalized to the population of the $v=4$, $J=1$  level, which is taken to be the reference level $r$, giving $N_r$ and $g_r$
We selected  $v=4$, $J=1$ to be the reference level because it is primarily excited by UV photons and its population is derived from the bright 4-2~O(3) line.
In many cases, there are multiple observed transitions arising from the same upper level (e.g. 1-0~S(1) and 1-0~Q(3)), providing independent measurements of $N_u$ for those upper states.

\subsection{Excitation Diagram of \htwo{} Level Populations} \label{sec:h2-level-pop}

\begin{figure*}
%\centering
\hspace{-1.2cm}
\includegraphics[width=1.09\textwidth]{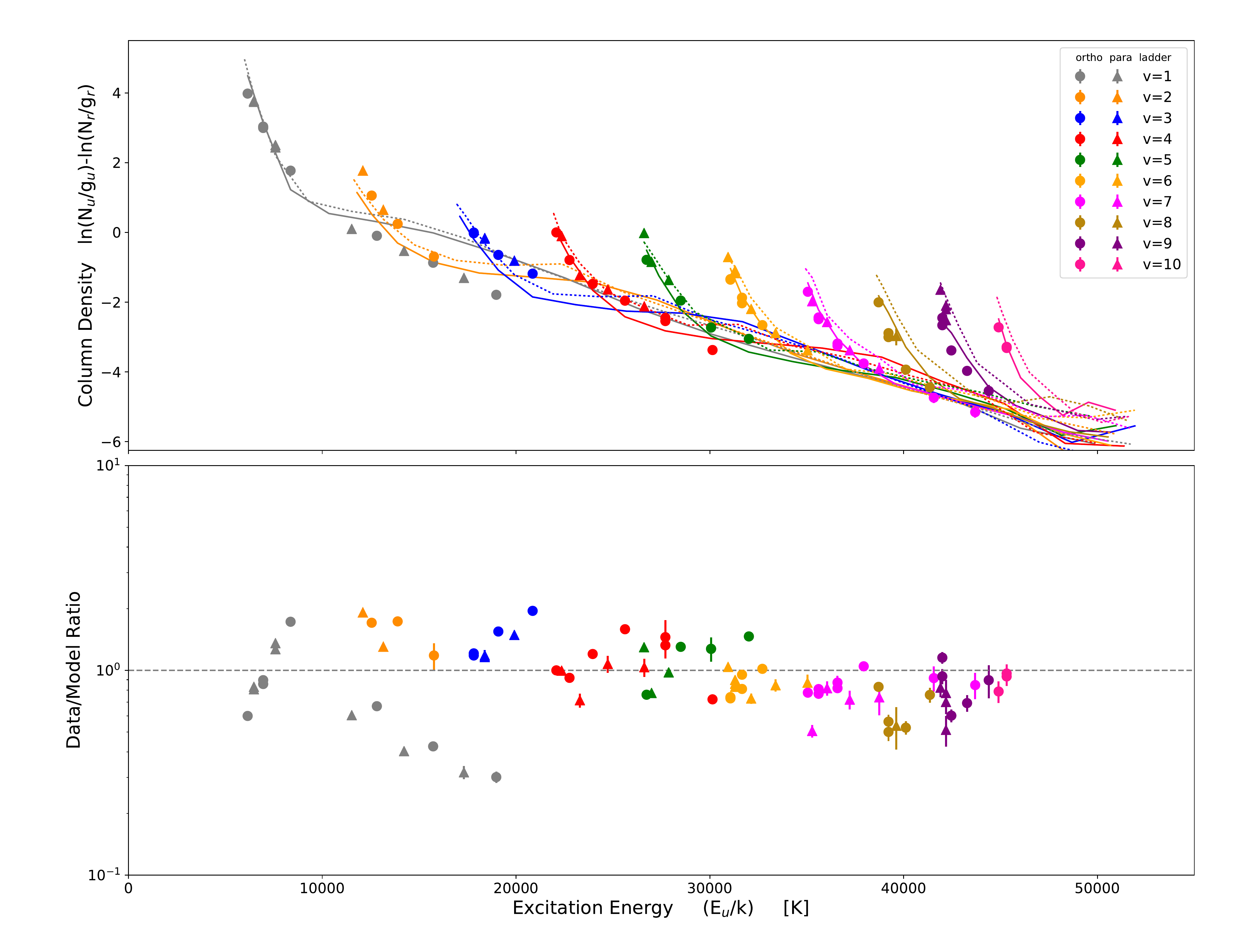}
\caption{
Top: excitation
diagram showing observed \htwo{} rovibrational level populations in the Orion Bar  as the data-points
vs. energy above the ground rovibrational ($v=0$, $J=0$) state.  Our
best-fit
constant density and temperature
($n_H=5\times10^3$~cm$^{-3}$ and $T=625$~K)
Cloudy model is shown
by
colored lines.  The error bars represent the $1 \sigma$ statistical uncertainty.  The solid lines are the fits for the ortho levels, and the dotted lines are fit for the para levels.   Both data and model are normalized to the reference 4-2~O(3) line ($v=4$, $J=1$ state), identified with $N_r$ and $g_r$ for the column density and quantum degeneracy respectively.  A purely isothermal gas would form a straight line on this diagram, while non-thermal mechanisms produce different patterns.  The  ``sawtooth'' pattern is characteristic of UV excitation.  
In dense gas, as seen here for the Orion Bar, collisions modify the level populations from the pure UV excited case.
Bottom: ratio
of the observed Orion Bar \htwo{} rovibrational level populations divided by the Cloudy model to show how well the model fits the data. The dashed line denotes a ratio of unity for which the data and model would be
in
perfect agreement.
}
\label{fig:model-best}
\end{figure*}

The top panel of Figure \ref{fig:model-best}
shows an excitation (or Boltzmann) diagram for the relative  \htwo{} rovibrational level column densities (or level populations) we observe in  the Orion Bar.
This diagram is a plot of the logarithmic column density of a transition's upper state $N_u$ divided by its quantum degeneracy $g_u$ vs. the excitation energy above the ground state ($v=0, J=0$), and is a convenient diagnostic tool for determining excitation mechanisms.

The spin of the two protons in \htwo{} can be either aligned or anti-aligned, forming two distinct spin isomers called ortho-\htwo{} (spins aligned) and para-\htwo{} (spins anti-aligned).  Since protons are fermions, the wave function of ortho-\htwo{} can only have odd values of
$J_u$, while
para-\htwo{} can only have even values of $J_u$.  In collisional equilibrium, the statistical weights for nuclear spin give an ortho-to-para ratio of
three.
The value of $g_u$ depends on the upper rotation state $J_u$ and whether the \htwo{} is ortho or para:
\begin{equation}
g_u^{\rm ortho} = 3 (2 J_u + 1),\ \ g_u^{\rm para} = 2 J _u + 1.
\end{equation}

\htwo{} that is primarily excited and de-excited by collisions (e.g. as in gas heated by a shock) has thermal rovibrational level populations.
In an isothermal region, the rovibrational level populations follow the Boltzmann distribution:
 \begin{equation} \label{eq:boltzmann}
\left({{N_u \over g_u} \Big/ {N_r \over g_r}}\right) \propto \exp\left(-{E_u \over k T}\right), \ \ \ln\left({{N_u \over g_u} \Big/ {N_r \over g_r}}\right) = -{E_u \over k T},
 \end{equation}
where $E_u$ is the energy above the  ground rovibrational state, $k$ is the Boltzmann constant, and
$T$ is the
kinetic temperature of the gas; in other
words,
the 
level populations follow a linear trend on an excitation diagram
with a slope inversely proportional to $T$.
If multiple temperature components are present, or there is a temperature gradient, the slope will flatten at higher excitation energies (e.g., \citealt{rosenthal00}).  This occurs because hotter gas dominates the excitation of states at the highest energies above ground, while cooler gas dominates the excitation of states at the lowest energies.

UV excitation of \htwo{} 
is a non-thermal process that leads to populations
that do
not show a monotonically decreasing trend for all the data-points on an excitation diagram, but instead follow a characteristic ``sawtooth''  pattern
(see Figure \ref{fig:model-best}).
The bulk of the \htwo{} in a PDR exists in the pure rotation $v=0$
states,
which lie at low enough excitation energies that collisions thermalize their level populations so that they reflect the underlying kinetic temperature of the gas.
UV excitation takes a small fraction of the underlying level populations of $J$ at $v=0$ and, in effect, transposes them to higher $v$.
The ``sawtooth'' pattern occurs because because quantum selection rules limit changes in $J$ during radiative transitions, but there are no such restrictions to changes in $v$.

One can fit straight lines to a series of rovibrational states of constant $v$ to derive a ``rotation temperature''  or across states of constant $J$ to derive a ``vibrational temperature,'' but one should be careful not to confuse these quantities with the actual kinetic temperature of the gas. 
Instead, they are shorthand for characterizing the relative level populations.
For UV excited \htwo{}, the level populations have high vibrational temperatures and lower rotation temperatures. 
While linear fits of these ladders (trends in constant $v$ or $J$) have been used in past studies of UV excited \htwo{}, they are not an ideal description for our information-rich data set, which probes up to high $J$ for many rotation ladders.   For example, some of the data-points in
the
$v=1$ rotation ladder deviate from a linear fit by up to $\sim 5$ orders of magnitude.
We therefore forgo the use of rotation or vibration temperatures in favor of comparing the rovibrational level populations we measure in the Orion Bar directly to values predicted by PDR models.

\section{Modeling and Interpretation} \label{sec:modeling}

\subsection{Simulating \htwo{} in 
the
Orion Bar
with
Cloudy} \label{sec:using-cloudy}

With IGRINS, we observe
87
NIR \htwo{} emission
lines,
which arise from 69 independent rovibrational states with excitation energies up to
$E_u/k = $50,000~K
above the ground ($v=0,~J=0$) state.
Our large dataset allows us to test our understanding of the physics in the Orion Bar by comparing the observed \htwo{} rovibrational level populations to model predictions.
For our models, we use version 13.03c of Cloudy\footnote{Cloudy: \url{http://nublado.org}.} \citep{cloudy13},
a one-dimensional plasma simulation code that solves for the physical conditions of a slab (or sphere) of gas irradiated by a photoionization source.   It includes detailed physics for radiative transfer through the gas, and the state of the constituent ions, atoms, molecules, and dust, and predicts the physical conditions of the gas and the emergent spectrum.
This version of Cloudy includes a fully
self-consistent
treatment of \htwo{} including the excited electronic and rovibrational states, radiative and collisional excitation, photodissociation, and reformation on dust grains \citep{shaw05}.

Collisions in dense gas, such as the Orion Bar,
can modify the \htwo{} rovibrational level populations \citep{sternberg89, burton90}.
For the Cloudy models, we have replaced the
\htwo{}--\hnaught{}
collision rate coefficients from \citet{wrathmall07} used by default in Cloudy 13.03c with  updated values from \citet{lique15}.  We use the default rates in Cloudy for 
\htwo{}--\htwo{}, \htwo{}--\hplus{}, and \htwo{}--He
collisions.   For collision rate coefficients
that
have no data (typically high $v$ and $J$), the ``g-bar approximation'' is used to estimate collision rate coefficients.
The g-bar approximation assumes that the rate coefficient for a collisionally induced transition is a function of that transition's change in energy  \citep{vanregemorter62, shaw05}.
We do not observe significant radial motion in the PV diagrams (see Figure \ref{fig:line-profiles}) expected from shock heated gas, consistent with the small line widths found by
\citet{burton90b}, \citet{parmar91}, and \citet{allers05}.
This confirms
that
we are observing collisionally modified UV exited \htwo{} as opposed to a combination of shocked and
low-density
UV excited \htwo{}. 
We compare the observed level populations to those for a grid of models with constant temperature and density, and to hydrostatic models of the Orion Bar.  In \sectionsymbol{}~\ref{sec:const-models} and \ref{sec:model-hydrostatic}, we present our
best-fit
model, along with a brief comparison to hydrostatic models to illustrate their advantages and disadvantages.

\subsection{Constant Temperature and Density Cloudy Models}\label{sec:const-models}

\subsubsection{Description of the Model Grid}

While constant temperature and density models do not properly capture the structure of the full Orion Bar PDR from the ionized zone to the cold molecular regions,  such simple models do reproduce the \htwo{} rovibrational level populations within the narrow \htwo{} emitting region.
It is possible that the temperature and density are nearly uniform across the narrow observed emitting region, explaining why these models provide good fits.
To explore the parameter space, we ran a grid of models with constant temperatures ranging
from $T = 200$  to  $800$~K  and
constant densities ranging
from $n_H=6.3\times10^2$ to $10^5$~cm$^{-3}$.
The gas turbulence and incident radiation field  (from the O7V star $\theta^1$~Ori~C) used in these models are taken from the Orion Bar Cloudy models by \citet{pellegrini09} and \citet{shaw09}.    This model grid allows us to explore the effects of different values of temperature and density on the \htwo{} rovibrational level populations.
Increasing the density and temperature increases the rate of collisions in the gas.
Each rovibrational level has a specific ``critical density'' for which the rate of collisional de-excitations equals the rate of radiative decays.

\subsubsection{Effects of Collisions}

Levels of low excitation energy, mainly the pure rotation states
($v=0$),
where the majority of the \htwo{} in a PDR lies, have low critical densities so their populations are primarily set by collisional excitation and de-excitation, which brings their populations into thermal equilibrium with the gas.  The kinetic temperature(s) of the gas sets the kinetic energy of the collisions, so increasing the temperature raises  the populations of the higher $J$ states.
For an isothermal region, the Boltzmann distribution (Equation \ref{eq:boltzmann}) describes these level populations.   If the gas is warm enough, collisions can excite some of the molecules to $v = $ 1, 2, and maybe 3.

The populations of levels with high critical densities, typically those with high excitation energies
at $v \geqslant 1$,
primarily depend on UV excitation and the subsequent radiative cascade set by the transition probabilities ($A_{ul}$) and other physical constants that are mostly invariant to external variables such as the UV radiation field intensity \citep{black87, sternberg88, sternberg89}.  
During UV excitation and the subsequent radiative cascade,
quantum selection rules allow transitions with all values of $\Delta v$ but restrict the value of $\Delta J$ to 0 or $\pm 2$.
In this sense, UV excitation crudely transposes the distribution of level populations for $J$ at $v=0$ to higher $v$. 
Since the gas kinetic temperature(s) sets level populations for $J$ at $v=0$, this shift of the $J$ level populations at $v=0$ to higher $v$ via UV excitation sets the relative column density of molecules at high $J$ for all $v$ values.  Making the gas warmer increases the relative column density of molecules at high $J$ for all $v$ and compresses each rotation ladder vertically in the logarithmic space of the excitation diagram
(Figure  \ref{fig:model-best})
while preserving the shape of a given rotation ladder (e.g., making the gas warmer vertically compresses the ``bent knee'' shape of the $v=1$ rotation ladder).

Increasing the temperature and/or density of the gas increases the rate of collisions, and this has differential effects on the populations of the lower energy $v=0$ and $1$  levels vs. the higher energy $v > 1$ levels.
As the collision rates increase,  the level populations in $v > 1$ become increasingly depressed by collisional de-excitation, while the $v=0$~and~$1$ levels are less depressed since they are at low enough energy to also be collisionally excited.
Increasing the temperature and/or density of the gas increases the rate of collisions,
and increases the suppression
of the level populations for $v > 1$.

Increasing the rate of collisions due to higher temperature and/or density has another effect on the level populations.  Collisionally induced transitions do not follow the same route to the ground level as the radiative cascade.
Radiative transitions favor low-$J$ and are limited by quantum selection rules ($\Delta J = 0$ or $\pm 2$), while collisionally induced transitions favor higher $J$ states and are not constrained by the same quantum rules.   This raises the population of the high-$J$ levels in a given rotation ladder and ``straightens'' the shape of the rotation and vibration ladders as seen on the excitation diagram.  This effect occurs simultaneously with, and at high density overwhelms, the compression of the rotation ladders caused by the redistribution of the collisionally excited $v=0$ levels to higher $v$ by UV excitation (e.g., the ``bent knee'' shape of the $v=1$ rotation ladder gets straightened into a monotonically decreasing trend).

\begin{figure*}
\hspace{-0.8cm}
\centering
\includegraphics[width=0.8\textwidth]{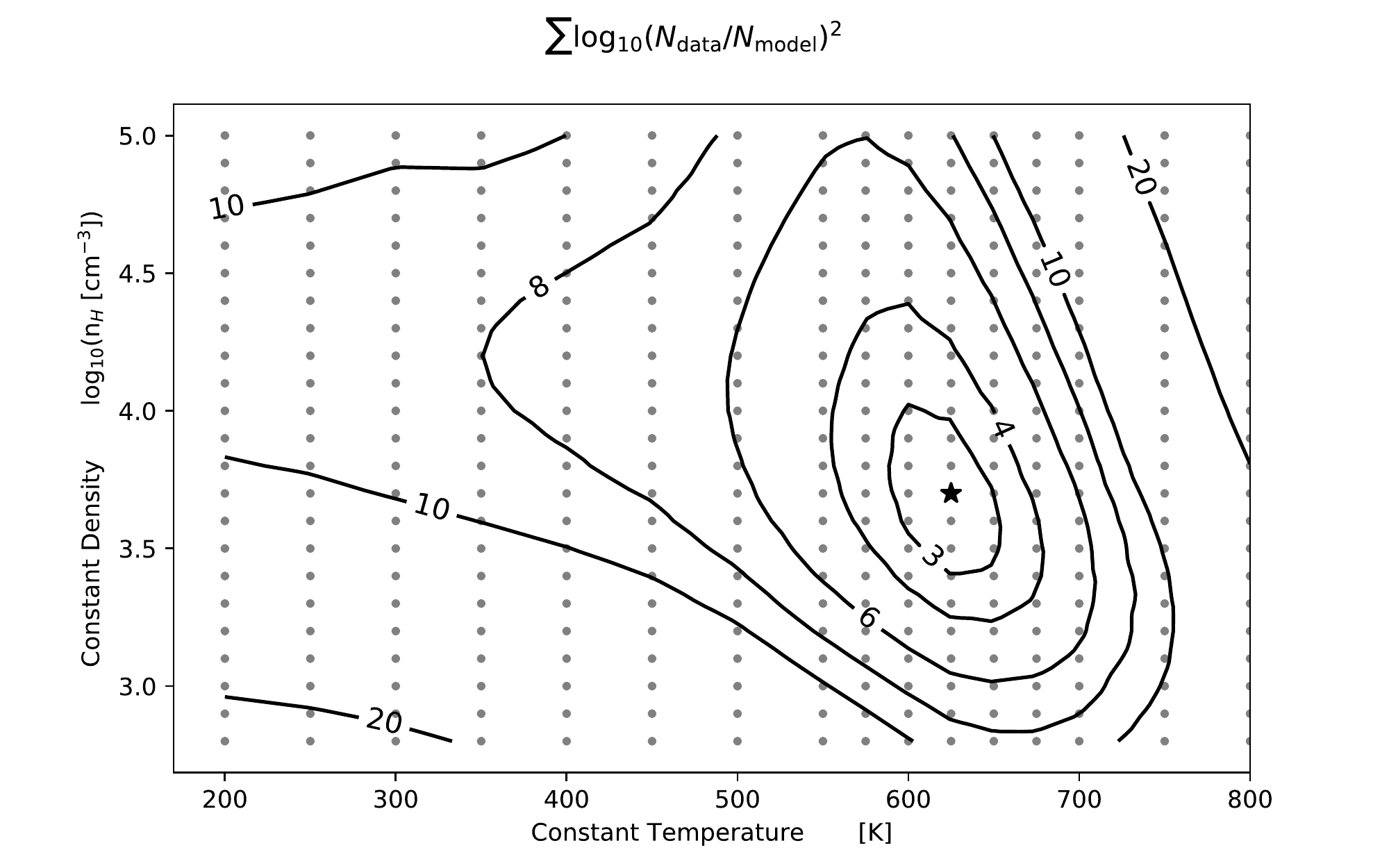}
\caption{
Contour plot of $\chi^2$ of the logarithm of the data-to-model ratio $\sum \log_{10}(N_{\rm data}/N_{\rm model})^2$ for determining how well a model fits the observed \htwo{} rovibrational level populations.  The constant temperature and constant density models on the grid range from
$T = 200$  to  $800$~K and $n_H=6.3\times10^2$ to $10^5$~cm$^{-3}$.
Each model is represented by a
grey point.
The
best-fit
model ($T=625$~K, $n_H=5\times10^3$~cm$^{-3}$) is represented by the black star.
}
\label{fig:model-grid}
\end{figure*}

\begin{table*}
\caption{
Best-fit
Cloudy Model Parameters}
\centering
\begin{tabular}{ll}
\hline
Parameter & Value   \\
\hline \hline
 Constant Temperature		& 625 			 K			\\
 Constant Density ($n_H$) 			& $5\times10^3$	 cm$^{-3}$ 	\\
 Turbulence		& 2				 km s$^{-1}$ $^a$	\\
 Abundances		& Orion$^b$						\\
 Grains			& Orion$^b$						\\
 Cosmic Ray Flux 	& Galactic Background$^b$			\\
Incident Radiation Field (O7V star $\theta^1$~Ori~C) & Kurucz Stellar  Atmosphere model, $T_{\rm eff} = 39700$ K$^a$		\\
No. of Ionizing Photons from  $\theta^1$~Ori~C	&	$Q(\mbox{H})=8.13 \times 10^{48}$ s$^{-1}$$^a$ \\
 Cloud Face Distance from  $\theta^1$~Ori~C & 0.114 pc$^a$		  \\
 Stopping Condition	&	$A_V = 14$ mag 			\\
No. of Iterations 			& 10							\\
\hline
\multicolumn{2}{l}{$^a$Parameters from Cloudy models of Orion Bar by \citet{pellegrini09} and \citet{shaw09}.} \\
\multicolumn{2}{l}{$^b$Stored prescription in Cloudy.}
\end{tabular}\label{tab:model-best}
\end{table*}

\subsubsection{Fitting the Model Grid}

With our model grid, we fit our observations of the Orion Bar and pinpoint the gas temperature and density by leveraging the effects of collisions on UV excited \htwo{}.
The main effect of higher density is to increase the collision rates.  Temperature affects both the collision rates and the thermal populations of the
$v=0$ ladder from which the relative level populations in $J$ are transposed to higher $v$ via UV excitation.
 Because of these dual effects of the temperature on the level populations, the model grid provides good leverage in fixing the gas temperature, while the density is less constrained.

We quantify the goodness of fit of the models to the data with a $\chi^2$ parameter of the logarithm of the data-to-model ratios $\sum \log_{10}(N_{\rm data}/N_{\rm model})^2$.  This gives all the data points equal weight regardless of the large dynamic range in the level populations.   Figure \ref{fig:model-grid} shows a contour plot of $\sum \log_{10}(N_{\rm data}/N_{\rm model})^2$ for the grid of constant density vs. constant temperature models, which are marked as dots.  
The
best-fit
model, marked with a star in Figure \ref{fig:model-grid}, 
has  $\sum \log_{10}(N_{\rm data}/N_{\rm model})^2 = 2.48$ with parameters
of $T=625$~K and $n_H=5\times10^3$~cm$^{-3}$.
Table \ref{tab:model-best} shows all input parameters for the
best-fit
model.  The other models in the grid have identical parameters except for temperature and/or density.
Figure \ref{fig:model-best} shows the Orion Bar data with the
best-fit
model's predicted level populations, and column 11 in Table \ref{tab:coldens} gives the ratios of the data to the
best-fit
model.  The level populations observed in the data and the
best-fit
model agree with each other within 0.5 dex. 

As expected, we find that only a narrow range of temperatures, between
600 and 650 K,
fits the data well.
This temperature range is consistent with the warm gas
($T = 250$ to $1000$~K)
observed via other species thought to coexist with rovibrationally excited \htwo{} in the Orion
Bar,
including excited pure rotation ($v=0$) lines of \htwo{} \citep{parmar91, allers05, shaw09},
ions such as C$^+$
\citep{tielens93, tauber94, wyrowski97},  and other excited molecules formed in the presence of excited \htwo{} \citep{nagy13}.  If the Orion Bar really does consist of a two phase medium with cooler dense clumps embedded in a warmer
low-density
medium (e.g., \citealt{burton90, parmar91, meixner93, andree-labsch14}), the \htwo{} emission we observed arises from the warmer
low-density
gas.   

The range of densities that fit the data is, again as expected, less well constrained than the temperature.  We get good fits
between $n_H = 2.5\times10^3$ to $10^4$~cm$^{-3}$,
which
are marginally
consistent with the values
of $n_H \geqslant 10^4$~cm$^{-3}$
reported by nearly all other measurements and estimates from excited \htwo{} and other species in the literature.  
If we assume pressure equilibrium where
$P/k\sim10^8$~cm$^{-3}$~K from \citet{goicoechea16}
and T = 600 to 650~K
from our model grid, we get
a density of $n_H\sim10^5$~cm$^{-3}$.  This is at least an order of magnitude greater than the densities of
$n_H = 2.5\times10^3$ to $10^4$~cm$^{-3}$
best fit by the model grid.
It is unclear why the best model fits have lower than expected
densities.
One possibility is that the UV excited \htwo{} emitting gas at our slit position actually has lower density than the majority of the gas in the Orion Bar, and previous studies of the Orion Bar used species such as the pure rotation lines of \htwo{} that trace the higher density gas.  Perhaps we are viewing the lowest density part of the cloud face, where self shielding is lowest and the UV radiation field interacting with the \htwo{} is strongest.  Another possibility is that some density-sensitive parameter(s) in the Cloudy models, such as the \htwo{} formation rate or the collisional rate coefficients, are overestimated or underestimated compared to their actual values.

The overall level populations across the different rotation ladders are well matched by the model, but the model over-predicts the observed level populations for high $J$ levels
in the $v=1$ ladder.
\citet{le16} found similar results
when fitting
models by \citet{draine96} to
their IGRINS observations of rovibrationally excited \htwo{} in the NGC 7023 PDR.
One possible explanation for this discrepancy is  ``formation pumping,'' where \htwo{} forms on dust grains in excited rovibrational states.  The distribution of rovibrational level populations for newly formed \htwo{} assumed in the models might be over-predicting the observed level populations at high $J$.
Cloudy assumes the prescription of \citet{takahashi2001} for formation pumping.
We ran a separate model grid with the formation pumping prescription of 
\citet{draine96} and
another set of grids with the formation pumping prescription set to thermal (Boltzmann) distributions with temperatures of 1500, 5000, 
10,000,
and
17,329\footnote{The default thermal formation pumping prescription in Cloudy has a temperature of
17,329 K,
corresponding to 1.5 eV or
one-third
of the energy released during the formation of an \htwo{} molecule, as described in \citet{lebourlot91}.} K.  We find that changing the formation pumping prescription in Cloudy does have a large effect on the predicted level populations at high $J$, but these alternate prescriptions do not provide better fits than the default \citet{takahashi2001} prescription.
The range of temperatures that best fit our data does not change significantly in any of these grids with alternative formation pumping prescriptions, but the range of
best-fit
densities
approaches $n_H=5\times10^4$~cm$^{-3}$ for
the Boltzmann distribution prescriptions
as the
temperature is lowered from
17,329
to 1500~K.
Since the high $J$ lines are sensitive to the adopted formation pumping prescription, using new formation pumping prescriptions or fine tuning existing prescriptions to fit the high $J$ levels might be an avenue for exploring formation pumping in future studies.

\subsection{Hydrostatic Models} \label{sec:model-hydrostatic}

We ran a suite of hydrostatic Cloudy models of the Orion Bar, based on the models from \citet{pellegrini09} and \citet{shaw09}.  These models were designed to simulate the full structure of the Orion Bar PDR and the \htwo{} emission.  
We ran these models with varying
cosmic-ray
fluxes, grain types, magnetic field strengths, temperature floors, and treatments for \htwo{} collisions.
While our
best-fit
constant density and constant temperature model fits the data 
well ($\sum \log_{10}(N_{\rm data}/N_{\rm model})^2 = 2.48$),
all the hydrostatic models provided poorer fits of $\sum \log_{10}(N_{\rm data}/N_{\rm model})^2>10$.

The main complication we find is that leaving the g-bar approximation on (as defined in \sectionsymbol{}~\ref{sec:using-cloudy}) yields unphysical rovibrational level populations, making it necessary to disable this feature.    Disabling the g-bar approximation means omitting some of the \htwo{} physics \citep{shaw05}.  This could introduce artificial effects between levels with
well-known
collision rate coefficients
(mainly levels with $v \leqslant 3$)
and those without, and it is unclear whether the predicted \htwo{} level populations for these models are physically meaningful.  
Turning the g-bar approximation off has a negligible effect on our constant temperature and density model grid fits.
Clearly,
there exists an interdependence between the collisional processes for \htwo{} and how the structure of the PDR is calculated in these hydrostatic models, that is less significant for the simpler constant temperature and density models. 
The hydrostatic model predictions for \htwo{} rovibrational level populations would greatly benefit from
well-known
collisional rate coefficients for transitions between high $v$ and $J$ states. 
New and improved collisional data for \htwo{} will ultimately give us a better understanding of PDR physics.

\section{Summary and Conclusions} \label{sec:conclusions}

We observed the Orion Bar PDR in a deep pointed observation with IGRINS at the 2.7 m telescope at the McDonald Observatory.  The instrument's high spectral resolution of
$R\sim$ 45,000
and broad wavelength coverage of the NIR H and K bands (1.45-2.45 $\mu$m) enables us to detect
87
\htwo{} rovibrational transition emission lines with S/N $> 3$.  
We extract the flux of each line with a robust weighting scheme and calculate the column density of \htwo{} for a total of 69 different rovibrational states, which have excitation energies up to
$E_u/k =$ 50,000 K
above the ground state ($v = 0$, $J = 0$).    The large range in rotational ($J$) levels, vibrational ($v$) levels, and excitation energy covered by the observed transitions allow us to perform a detailed study of the excitation of \htwo{} within the Orion Bar PDR.   We compare the observed rovibrational level populations to predictions from one-dimensional Cloudy 13.03c \citep{cloudy13} models. 

As a result of our analysis, we find
the following.
\begin{enumerate}
\item The spectral resolution of IGRINS
($R\sim$ 45,000)
is high enough that the wavelengths for the \htwo{} rovibrational transitions calculated from the experimentally determined \htwo{} ground electronic state rovibrational energy levels in \citet{dabrowski1984} were found to differ from the observed wavelengths by up to $10^{-4}$~$\mu$m.  New wavelengths calculated from the theoretical energy levels in \citet{komasa11} provide almost an order of magnitude improvement
in the agreement between observed and calculated line wavelengths, with the majority of the lines agreeing to within the uncertainty of our wavelength calibration
($< 6 \times 10^{-6}$~$\mu$m).
\item
The line-of-sight extinction towards the \htwo{} emitting region is $A_V = 8.25$ or  $A_K = 0.96$,
as measured from line pairs arising from common upper states.
\item Constant temperature and density Cloudy models provide a better fit to the IGRINS \htwo{} data than the hydrostatic models of \citet{shaw09} and \citet{pellegrini09}, which explicitly solve for the structure throughout the PDR and have nearly constant pressure.  This could be due to the fact that 
the $v \geqslant 1$ transitions
we observe in the Orion Bar arise from a relatively narrow zone of the overall PDR
structure.
%where the temperature and density are nearly uniform. 
Another possible explanation for the poorer fit of the hydrostatic models is that this results from disabling the g-bar approximation for the collisional rate coefficients of the high-$v$ levels \citep{vanregemorter62, shaw05}, which may omit physical effects important in determining the level populations.
\item The model grid, combined with the large number of rovibrational levels we probe, constrains the temperature for the
observed \htwo{} emitting region 
to 600 to 650 K,
consistent with earlier findings.  The
best-fit
model gives a temperature of
625~K.
\item The model grid constrains the density to
$n_H =  2.5\times10^3$ to $10^4$~cm$^{-3}$,
with the
best-fit
model giving $n_H = 5 \times 10^3$~cm$^{-3}$, which is
marginally
lower than most values in the literature.
The reason may
be either that this emission arises predominantly in the
lower density
inter-clump region of a two-component clumpy medium (with which our density is marginally consistent), or that one or more of the assumed parameters in the Cloudy models are sensitive to density and their values are
over-
or underestimated.
\end{enumerate}

\acknowledgments
This work used the Immersion Grating INfrared Spectrometer (IGRINS) that was  developed under a collaboration between the University of Texas at Austin and the  Korea Astronomy and Space Science Institute (KASI) with the financial support of  the US National Science Foundation under grant
 AST-1229522 to
the University of Texas at Austin, and of the Korean GMT Project of KASI.
This paper includes data taken at The McDonald Observatory of The University of Texas at Austin.
We acknowledge the Cambridge Astronomical Survey Unit and WFCAM Science Archive for making available data that were used for the finder charts in Figure \ref{fig:full_finder}.
We would like to acknowledge Gary Ferland for helpful discussions on the Cloudy modeling, and
Evelyne Roueff for pointing out the theoretical \htwo{} ground electronic state rovibrational energy levels in \citet{komasa11} from which we derive improved line wavelengths for the rovibrational transitions (\sectionsymbol{}~\ref{sec:wavelengths}).

\bibliography{../../literature/literature.bib} %Show references

\end{document}